\documentclass{article}

\usepackage{amsmath}  
\usepackage[english]{babel}
\usepackage[utf8]{inputenc}
\usepackage{amssymb}
\usepackage{graphicx}
\usepackage{authblk}
\usepackage{caption}
\usepackage{multirow}
\usepackage{amscd}
\usepackage{geometry}
\usepackage{a4wide}
\usepackage[usenames]{color}
\usepackage{etaremune} 
\usepackage[hidelinks]{hyperref}
\usepackage{amsfonts}
\usepackage{ifthen}
\usepackage{color}
\usepackage{amscd}
\usepackage{latexsym}
\usepackage{mathrsfs}

\newcommand{\bd}{\begin{definition}}
	\newcommand{\ed}{\end{definition}}
\newcommand{\bt}{\begin{theorem}}
	\newcommand{\et}{\end{theorem}}
\newcommand{\bi}{\begin{itemize}}
	\newcommand{\ei}{\end{itemize}}
\newcommand{\ben}{\begin{enumerate}}
	\newcommand{\een}{\end{enumerate}}
\newcommand{\beq}{\begin{equation}}
	\newcommand{\eeq}{\end{equation}}

\newcommand{\R}{\mbox{$ \mathbb{R}  $}}

\newcommand{\hr}{\mbox{$ \mathcal{H}(2,\mathbb R) $}}
\newcommand{\spinf}{\mbox{$ \mathbb R \oplus \R^2 $}}

\DeclareMathOperator{\Tr}{Tr}

\newtheorem{definition}{Def.}[section]
\newtheorem{theorem}{Theorem}[section]
\newtheorem{proposition}{Proposition}[section]

\newtheorem{corollary}{Corollary}[section]

\def \proof{\noindent{\it Proof. \;}  \ignorespaces}
\def \qed{ \hfill $\Box$ \\}

\def \proof{\noindent{\it Proof}.  \ignorespaces}
\def \qed{ \hfill $\Box$ \\}

\begin{document}
	
	\title{Quantum measurement and color perception: theory and applications}
	\author[1]{Michel Berthier\thanks{michel.berthier@univ-lr.fr}}
	\author[2]{Edoardo Provenzi\thanks{edoardo.provenzi@math.u-bordeaux.fr}}
	\affil[1]{Laboratoire MIA, Pôle Sciences et Technologie, Université de La Rochelle, 23 Avenue A. Einstein, BP 33060, 17031 La Rochelle Cedex, France}
	\affil[2]{Université de Bordeaux, CNRS, Bordeaux INP, IMB, UMR 5251. F-33400, 351 Cours de la Libération}

	\renewcommand\Authands{ and }
	\date{}

\maketitle

\begin{abstract}
In this paper we make a systematic use of the quantum measurement theory to describe perceived colors and analyze some of their fundamental properties. After motivating the naturalness of the quantum measurement approach in the mathematical framework of the color perception theory proposed by the authors in previous papers, we show how to obtain several results. Among our  theoretical outcomes, we mention  the possibility to confine the color cone to a finite-volume color solid and the link between post-measurement state changes, Lorentz boosts and the Einstein-Poincaré relativistic addition law. We apply these results to obtain a chromatic match equation that emphasizes the importance of the Hilbert-Klein metric on the unit disk and we also present a quantitative description of Hunt's effect.
\end{abstract}

\section{Introduction}
The main aim of this contribution is to integrate in a systematic way some important elements of  quantum measurement theory into the mathematical framework of color perception proposed in the papers \cite{BerthierProvenzi:19,Berthier:2020,Berthier:2021JofImaging,Berthier:21JMIV,Berthier:21JMP}. This framework describes a quantum-relativistic theory of color perception, based on a totally different paradigm with respect to the one assumed in the classical colorimetry developed by CIE (Commission Internationale de l'Éclairage), which, essentially, can be resumed as a metameric reduction of the space of physical color stimuli, see for instance \cite{Krantz:75,Cohen:85,Dubois:09} for  mathematically-oriented descriptions of this methodology.

Since the direct link between light stimuli and perceived colors is lost during the metameric reduction and since the human visual system elaboration separates these two concepts even further, we have decided not to consider physical color stimuli and to base our model solely on well-known empirical evidences about color perception. As we will detail later, this strategy has been already followed by important scientists. 

These empirical color perception laws cannot be investigated without making reference to an experimental environment and to an observing apparatus, as incisively pointed out by the logician B. Russell in \cite{Russell:16}: `\textit{When, in ordinary life, we speak of the colour of the table, we only mean the sort of colour which it will seem to have to a normal spectator from an ordinary point of view under usual conditions of light. But the other colours which appear under other conditions have just as good a right to be considered real; and therefore, to avoid favoritism, we are compelled to deny that, in itself, the table has any one particular colour}'. 

The need to understand empirical evidences dependent on the experimental environment and on the observing conditions is one of the motivations that triggered the mathematical formalization of quantum mechanics and, as we will recall in section \ref{sec:recap}, the formalism of quantum theories reveals to be perfectly fit to model color perception too. 

The novel contributions of this paper descend from the mathematical investigation of the consequences of the very act of perceiving a color, modeled as a nondestructive quantum measurement performed by the human visual system, extending an idea pioneered in \cite{Berthier:2020}.

Our study starts with the introduction of the fundamental concepts of positive operator-valued measure, effect and unsharp observable in section \ref{sec:quantumeasurement}\ref{subsec:POVMandeffects}. The relevance of these concepts in color perception will be first discussed in section \ref{sec:quantumeasurement}\ref{subsec:colorsolid} by showing that effects provide an elegant and natural solution to the long-lasting problem of confining the infinite conical color space into a finite volume, in perfect agreement with the color solid advocated by Ostwald and deValois.

In section \ref{sec:colormeasurelativistic}\ref{subsec:Luders} we integrate other fundamental tools of quantum measurement theory, such as Lüders operations, generalized density matrices and Kraus operators, into the quantum theory of color perception. In the two following subsections, \ref{sec:colormeasurelativistic}\ref{subsec:boosts} and \ref{sec:colormeasurelativistic}\ref{subsec:EP} respectively, we show that the Lüders operation that describes the state transformation associated to an effect is proportional to a Lorentz boost and that the post-measurement state is parameterized by a vector obtained as the Einstein-Poincaré relativistic addition of the so-called chromatic vectors associated to the effect and the initial state. This reveals the tight link between color perception and relativity.

Section \ref{sec:chromaticmatch} will be dedicated to colorimetric applications of the previous results. In particular, in section \ref{sec:chromaticmatch}\ref{subsec:chromaticmatchingeq} we will translate chromatic matching into an equation between two post-measurement states. This result will then be used in section \ref{sec:chromaticmatch}\ref{subsec:HilbertKlein} to single out the Hilbert-Klein metric on the unit disk as a pertinent perceptual chromatic distance, extending and completing the partial result on the same topic obtained in \cite{Berthier:21JMP}.

The last application is devoted to Hunt's perceptual colorimetric effect which, from our point of view, is a very relevant illustration of the effectiveness of quantum information to model the environmental and observational dependence of color perception. We will show how it is possible to quantify the variation of  colorfulness in terms of variation of the perceptual counterpart of luminance. 

Finally, in section \ref{sec:conclusions} we will discuss some  perspectives for future investigations.

\section{Prerequisites about a quantum-like color perception model}\label{sec:recap}
The whole content of this section is scattered in the contributions  \cite{BerthierProvenzi:19,Berthier:2020,Berthier:2021JofImaging,Berthier:21JMIV,Berthier:21JMP}, however, for the sake of self-consistency of the present paper and also to have at disposal the concepts and the nomenclature for the development of our original contributions, it is essential to recap the most important results of the papers quoted above. The reader interested in a more detailed discussion of the basic colorimetric principles underlying our work can consult \cite{Berthier:21JMP,Provenzi:20}.

Schrödinger's axioms \cite{Schroedinger:20} efficiently resumed previous findings by Newton, Maxwell, Young, von Helmholtz and Grassmann about the color sensation induced by a light stimulus observed in isolation from the rest of a visual scene, a very restrictive (yet standard) condition in colorimetry.

Schrödinger's axioms establish that the set of perceptual colors is not merely a collection of sensations, but it possesses the structure of a 3-dimensional regular convex cone. As proven by Resnikoff in \cite{Resnikoff:74}, see also \cite{Provenzi:20}, if one adds to Schrödinger's description the property of homogeneity, then the geometry of the space describing color sensations registered by trichromatic human observers can be described only as follows: $\mathcal C_1=\R^+\times \R^+\times \R^+$ or $\mathcal C_2=\R^+ \times \textbf{H}$, where $\textbf{H}$ is a 2-dimensional hyperbolic space with constant negative curvature $-1$, in Resnikoff's paper $\textbf{H}\equiv \text{SL}(2,\R)/\text{SO}(2)$. 

When equipped with the so-called Helmoltz-Stiles metric, i.e. $ds_{HS}^2=\alpha(dx/x)^2+\beta(dy/y)^2+\gamma(dz/z)^2$, $\alpha,\beta,\gamma\in \R^+$, the flat color space $\mathcal C_1$ coincides with the classical colorimetric space used by the CIE, which has  proven unfit to describe perceptual features, see e.g. \cite{Provenzi:20,Berthier:21JMIV,Fairchild:13}. Instead, the color space $\mathcal C_2$, investigated in the series of papers \cite{BerthierProvenzi:19,Berthier:2020,Berthier:2021JofImaging,Berthier:21JMIV,Berthier:21JMP} did not show any inconsistency.

Crucially, $\mathcal C_1$ and $\mathcal C_2$ happen to be the positive cones of the only two non-isomorphic formally real Jordan algebras\footnote{The positive cone of a Jordan algebra is the interior of the set of its squares. We refer to the excellent references \cite{Baez:12,Faraut:94,Mccrimmon:1978} for details about Jordan algebras.} (FRJA from now on) of real dimension 3. Thus, Schrödinger's and Resnikoff's axioms can be reduced to the following assumption called \textit{Trichromacy axiom} in \cite{Berthier:2020}: \textit{the perceptual color space $\mathcal C$ is the positive cone of a formally real Jordan algebra of real dimension 3.}

\medskip
$\mathcal C_1$ is the positive cone of the FRJA $\R\oplus \R \oplus \R$, which is associative and not simple, $\mathcal C_2$ is the positive cone of either the FRJA $\hr$, given by real symmetric $2\times 2$ matrices, or $\spinf$, the so-called spin factor, both endowed with a suitable Jordan product. $\hr$ and $\spinf$ are non-associative, simple and naturally isomorphic as Jordan algebras via the following isomorphism:
\begin{equation}\label{eq:isoH2}
	\begin{array}{cccl}
		\chi: & \mathcal H(2,\mathbb R)  & \stackrel{\sim}{\longrightarrow} & \mathbb R\oplus \mathbb R^2  \\
		& \begin{pmatrix}
			\alpha + v_1 & v_2 \\
			v_2 & \alpha - v_1
		\end{pmatrix}  & \longmapsto & \begin{pmatrix}
			\alpha \\ {\bf v}
		\end{pmatrix}, \qquad \alpha\in \R, \; {\bf v}=(v_1,v_2)^t\in \R^2,
	\end{array}
\end{equation}
which induces the following isomorphism between their positive cones:
\beq\label{eq:cones2}
\begin{array}{ccc}
	\mathcal C(\mathbb R \oplus \mathbb R^2) & = & \mathcal L^+ \\
	\rotatebox{90}{$\cong$} &    & \rotatebox{90}{$\cong$} \\
	{\mathcal C(\mathcal H(2,\mathbb R))}   & = & {\mathcal H^{+}(2,\mathbb R)}  
\end{array},
\eeq 
where $\mathcal L^+$  is the \textit{future lightcone}, i.e. the subset of the 3-dimensional Minkowski space $\mathcal M=\R^{1,2}$ given by the elements $a=(\alpha,\textbf{v})^t$ such that $\alpha >0$ and the squared Minkowski norm of $a$ is positive, i.e. $\|(\alpha, \textbf{v})\|^2_{\mathcal M}:=\alpha^2 - \|\textbf{v}\|^2>0$, $\|\; \|$ being the Euclidean norm. $\mathcal H^{+}(2,\mathbb R)$ is the set of positive-definite $2\times 2$ real matrices, i.e. $A\in \mathcal H(2,\mathbb R)$ such that $\textbf{v}^tA\textbf{v}>0$ for all $\textbf{v}\in \R^2\setminus \{\textbf{0}\}$.

The positive cone of a FRJA $\mathcal A$ is always self-dual, i.e. it satisfies the following property:
\beq\label{eq:dualcone}
\mathcal C(\mathcal A)=\mathcal C'(\mathcal A):=\{a\in \mathcal A \; : \; \forall b\in \mathcal C, \; \langle a,b\rangle >0  \}\cong \{ \omega \in \mathcal A^* \; : \; \forall b\in \mathcal C, \; \omega(b) >0 \}=:\mathcal C^*(\mathcal A),
\eeq 
where, once defined $L_{a\circ b}(c):=(a\circ b)\circ c$,  $\langle a,b\rangle := \text{Tr}(L_{a\circ b})$ is a well-defined inner product on $\mathcal A$ and the isomorphism between $\mathcal C'(\mathcal A)$ and $\mathcal C^*(\mathcal A)$ is guaranteed by Riesz's  representation theorem.

The domain of positivity of the Jordan algebras $\hr$ and $\spinf$, i.e. the set of their squares, coincides with the closure of their positive cones:
\beq\label{eq:cones1}
\begin{array}{ccc}
	\overline{\mathcal C}(\mathbb R \oplus \mathbb R^2) & = & \overline{\mathcal L^+}\\
	\rotatebox{90}{$\cong$} &    & \rotatebox{90}{$\cong$} \\
	\overline{\mathcal C}(\mathcal H(2,\mathbb R)) & = & {\overline{\mathcal H^+}(2,\mathbb R)}   
\end{array},
\eeq 
where $\overline{\mathcal L^+}$ is defined by $\alpha\ge 0$ and $\|(\alpha,\textbf{v})\|^2_{\mathcal M}\ge 0$, i.e. $\alpha^2 - \|\textbf{v}\|^2\ge 0$. Instead, $\overline{\mathcal H^+}(2,\mathbb R)$ is the set of positive semi-definite $2\times 2$ matrices, i.e. $A\in \hr$ such that $\textbf{v}^tA\textbf{v}\ge 0$ for all $\textbf{v}\in \R^2$.

In this paper we will deal only with the space $\mathcal C_2$ because of its accordance with perceptual observations, its rich geometrical properties and, last but not least, because it offers the possibility to build a quantum-like theory of color perception. 

This point is crucial for the comprehension of the rest of the paper. Russel's words reported in the introduction clearly underline the central role of both the preparation of a visual scene and the color measurement process in the perceptual description of color. This is clearly a philosophical view very close to that of quantum mechanics, in which it does not even make sense to discuss physical properties without specifying the experimental conditions of the system and how this one interacts with the measurement apparatus. The consequences that a quantum-like measurement of color entail on our theory will be analyzed starting from section \ref{sec:quantumeasurement}.

Non-associative Jordan algebras have been proven to provide a perfectly valid framework to develop quantum theories in the pioneering paper \cite{Jordan:34}, in the sense that their algebraic description of states and observables is equivalent to the density matrix formalism that can be built starting from the ordinary Hilbert space formulation \cite{Townsend:85,Emch:2009}. Non-commutativity of Hermitian operators on a Hilbert space is replaced by non-associativity in the Jordan framework, this is essential to preserve the core of quantum theories, i.e. the existence of uncertainty relations, which cannot appear if the Jordan algebra of observables is both commutative and associative. 

It is with these premises that, in the papers \cite{BerthierProvenzi:19,Berthier:2020,Berthier:2021JofImaging,Berthier:21JMIV,Berthier:21JMP} a quantum-like theory of color perception has been developed. We underline that this theory does not consist simply in a modification of the language or the mathematical tools used to analyze the classical colorimetric theory, but it is a genuine change of paradigm: we no more have a color description in terms of three coordinates belonging to a flat color space, but a quantum-like theory of \textit{color states and observables in duality with each other} and where, as we will point out in section \ref{sec:quantumeasurement}, \textit{perceived colors are inextricably associated with measurements, described by quantum effects}.

In order to carry on our theory of color perception unambiguously, we formalize the concepts just discussed through the following fundamental and interconnected definitions, clearly inspired by the quantum axiomatic.

\begin{definition}
	A visual scene is an environment that allows human observers to perform measurements through their visual system and to register the outcomes as perceived colors.
\end{definition}

\begin{definition}
	A perceptual observable, or simply an observable, is defined by an \textit{equivalence class of measurements} performed on a visual scene, where two measurements are equivalent if, no matter how they are carried out, they lead to the registration of the same set of outcomes. 
\end{definition} 

As a direct consequence of the trichromacy axiom, the algebra of observables $\mathcal A$ of a quantum-like theory of color perception  is either $\hr$ or $\spinf$. 

\begin{definition}
	A perceptual state, or simply a state, is an equivalence class of preparations of a visual scene, the equivalence being that, independently of preparation protocol, the measurements outcomes are the same.
\end{definition} 

In the algebraic formulation of quantum mechanics states are described by density matrices, i.e. unit-trace positive semi-definite matrices. The \textit{state vectors}  $\textbf{v}_ {\bf s}=(s_1,s_2)^t$ of the unit disk $\mathcal D$ parameterize each density matrix $\rho_ {\bf s}$ representing perceptual chromatic states, in fact  \cite{Berthier:2020}:
\beq\label{eq:dmat}
\mathcal S(\mathcal H(2,\R)) = \{\rho_ {\bf s} \in \overline{\mathcal H^+}(2,\R), \; \text{Tr}(\rho_ {\bf s})=1 \} = \left\{ \rho_ {\bf s}\equiv \frac{1}{2}\begin{pmatrix}
	1+s_1 & s_2 \\ s_2 & 1-s_1
\end{pmatrix}, \;  \|\textbf{v}_ {\bf s}\|\le 1 \right\},
\eeq 
and also, as a consequence of (\ref{eq:isoH2}),
\begin{equation}\label{eq:chirhos}
	\mathcal S(\mathcal \spinf) =\chi( \mathcal S(\mathcal H(2,\R)))= \left\{ \chi(\rho_ {\bf s})=\frac{1}{2}\begin{pmatrix}
		1 \\ {\bf v}_{\bf s}
	\end{pmatrix}, \; \Vert{\bf v}_{\bf s}\Vert\leq 1 \right\}. 
\end{equation}
Polar coordinates are the most natural ones in $\mathcal D$ and provide this alternative parameterization of the generic density matrix:
\begin{equation}\label{eq:rhortheta}
	\rho_ {\bf s}(r,\vartheta)=\frac{1}{2}\begin{pmatrix}
		1+r\cos\vartheta&r\sin\vartheta\\r\sin\vartheta & 1-r\cos\vartheta \end{pmatrix}, \qquad r\in [0,1], \; \vartheta\in [0,2\pi).
\end{equation}
States can be either \textit{mixed} or \textit{pure}, accordingly to the fact that they can be written as a convex combination of other states or not, respectively. An extremely useful quantitative measure of purity is provided by the \textit{von Neumann entropy}: 
\beq 
S(\rho_ {\bf s})=-\Tr(\rho_ {\bf s} \log_2 \rho_ {\bf s}).
\eeq 
$S(\rho_ {\bf s})$ represents \textit{the expectation of information gain on a quantum system after a measurement}. If $S(\rho_ {\bf s})=0$, then no information gain can be achieved after a measurement, this can happen only if all the available information about the system in that state $\rho_ {\bf s}$ is known, in this case $\rho_ {\bf s}$ is called a pure state. The opposite situation is when $S(\rho_ {\bf s})$ is maximal, clearly corresponding to the least available information about the system, because every measurement will provide a maximal expectation of information gain.

A straightforward calculation shows that the von Neumann entropy in our case is the concave radial function expressed by:
\beq
S(r)=-\frac{1-r}{2}\log_2\left(\frac{1-r}{2}\right)-\frac{1+r}{2}\log_2\left(\frac{1+r}{2}\right), 
\eeq 
for $0\le r<1$ and $S(1)=0$, independently of $\vartheta \in [0,2\pi)$. The state of maximal von Neumann entropy corresponds to $r=0$, the center of the unit disk in $\R^2$, and it is represented by the density matrix $\rho_{\bf0}=Id_2/2$,  or equivalently by $\chi(\rho_{\bf0})=\frac{1}{2}(1,{\bf0})$,  and $S(0)=1$. 

Instead, pure states are parameterized by the points of the border of $\mathcal D$:
\beq
\mathcal {PS}(\mathcal H(2,\R)) = \{\rho_ {\bf s} \in \mathcal S(\mathcal H(2,\R)) , \; \text{Tr}(\rho_ {\bf s}^2)=1 \} = \left\{ \rho_ {\bf s}\equiv \frac{1}{2}\begin{pmatrix}
	1+s_1 & s_2 \\ s_2 & 1-s_1
\end{pmatrix}, \;  \|\textbf{v}_ {\bf s}\|= 1 \right\}
\eeq 
or, equivalently,
\begin{equation}
	\mathcal {PS}(\mathcal \spinf) =\chi( \mathcal {PS}(\mathcal H(2,\R)))= \left\{ \chi(\rho_ {\bf s})=\frac{1}{2}\begin{pmatrix}
		1 \\ {\bf v}_{\bf s}
	\end{pmatrix}, \; \Vert{\bf v}_{\bf s}\Vert = 1 \right\}. 
\end{equation}
If we define $\sigma_i=\chi^{-1}(e_i)$ for all $i=0,1,2$, where $(e_0,e_1,e_2)$ is the canonical basis of $\R^{1,2}$, then we obtain $\sigma_0= Id_2$ and
\begin{equation}
	\sigma_1=\left(\begin{array}{cc}1 & 0 \\0 & -1\end{array}\right),\ \ \sigma_2=\left(\begin{array}{cc}0 & 1 \\1 & 0\end{array}\right),
\end{equation}
the two real Pauli matrices. $(\sigma_0,\sigma_1,\sigma_2)$ is an orthogonal basis for $\mathcal H(2,\R)$, in fact it satisfies:
\beq  
\sigma_i \cdot \sigma_j := \Tr(\sigma_i\sigma_j)=2\delta_{ij}, \qquad i,j=0,1,2.
\eeq 
As a direct consequence, the generic density matrix of $ \mathcal S(\mathcal H(2,\R))$ can be written as follows:
$$
\rho_ {\bf s}= {1\over 2}(Id_2+s_1\sigma_1 + s_2\sigma_2)=\rho_{\bf0} +{1\over 2}(s_1\sigma_1 + s_2\sigma_2) , \qquad \|\textbf{v}_ {\bf s}\|\le 1.
$$
By a direct algebraic computation one can obtain the following crucial formula \cite{Berthier:2020}:
\beq\label{eq:Heringopp}
\begin{split}
	\rho(r,\vartheta) &= \rho_{\bf 0} +\frac{r\cos \vartheta}{2}\sigma_1 + \frac{r\sin \vartheta}{2}\sigma_2 \\
	& = \rho_{\bf 0}+{r\cos\vartheta\over 2}\left[\rho(1,0)-\rho(1,\pi)\right]+{r\sin\vartheta\over 2}\left[\rho\left(1,{\pi\over 2}\right)-\rho\left(1,{3\pi\over 2}\right)\right] \, ,
\end{split}
\eeq 
for all $r\in [0,1]$, $\vartheta \in [0,2\pi)$, with 
\begin{equation}\label{eq:rho0}
	\rho_{\bf 0} ={1\over 4}\rho(1,0)+{1\over 4}\rho(1,\pi)+{1\over 4}\rho\left(1,{\pi\over 2}\right)+{1\over 4}\rho\left(1,{3\pi\over 2}\right).
\end{equation}
In \cite{Berthier:2020,Berthier:2021JofImaging} the following quite natural identifications have been performed:
\begin{itemize}
	\item[-] the \textit{saturation} associated to a chromatic state $\rho(r,\vartheta)$ is represented by the function:
	\beq\label{eq:sat}
	\Sigma(r)=1-S(r)=\frac{1}{2}\log_2(1-r^2)+\frac{r}{2}\log_2\left(\frac{1+r}{1-r}\right), \quad r\in [0,1)
	\eeq 
	and $\Sigma(1)=1$, independently of the value of coordinate $\vartheta$;
	\item[-] the pure states $\rho(1,\vartheta)$ are \textit{maximally saturated chromatic states}, or \textit{pure hues};
	\item[-] for all $\vartheta\in [0,\pi)$, the antipodal pure states $\rho(1,\vartheta)$ and $\rho(1,\vartheta+ \pi)$ describe \textit{opponent chromatic states};
	\item[-] $\rho_{\bf 0}$ represents the \textit{achromatic state} because it is the mixed state built by the perfect balance between two opponent chromatic states. 
\end{itemize} 
Notice that, since $\rho_{\bf 0}$ is labeled by $r=0$, the saturation associated to the achromatic state is $\Sigma(0)=0$, while pure states, labeled by $r=1$, are associated to a saturation $\Sigma(1)=1$. Thus, definition \eqref{eq:sat} is perfectly coherent with the common colorimetric concept of saturation, with the advantage of having a precise analytical form.

Thanks to these identifications, we can easily interpret formula \eqref{eq:Heringopp}: a mixed chromatic state $\rho(r,\vartheta)$, with $(r,\vartheta)\in (0,1) \times [0,2\pi)$, can be seen as the result of the contribution of the achromatic state $\rho_{\bf 0}$ plus two pure chromatic states oppositions. Following Hering \cite{Hering:1878}, we can identify the opposition encoded by the Pauli matrix $\sigma_1$ with red vs. green (R-G) and the one encoded by the Pauli matrix $\sigma_2$ with yellow vs. blue (Y-B). 

We remark that, while the strengths of the opponent contributions are dependent on $(r,\vartheta)$, $\rho_{\bf 0}$ in formula \eqref{eq:Heringopp} appears as a sort of `offset state', independent of $(r,\vartheta)$.  This description is perfectly in line with Hering's theory, as underlined very well by D. Hubel in \cite{Hubel:95}: \textit{`in Hering's theory the black and white process required a spatial comparison $[\dots]$, whereas his yellow-blue and red-green processes represent something occurring in one particular place in the visual field, without regard to the surround}'. 

Using a vocabulary closer to our model, and recalling that we are modeling color perception in isolated conditions, we can re-interpret Hubel's words by saying that formula \eqref{eq:Heringopp} shows that \textit{the two degrees of chromatic opposition constitute an intrinsic characteristic of a perceptual color state} isolated from its environment, or, equivalently, that an isolated part of a visual scene can be prepared for the measurement of a couple of chromatic opponencies. On the contrary, since $\rho_{\bf 0}$ is common to all mixed states, \textit{the achromatic state is an extrinsic feature of a perceptual color state}, or, equivalently, it is not possible to prepare an isolated part of a visual scene for the measurement of its achromatic component, a comparison with the surround is essential. 

Regarding pure states $\rho(1,\vartheta)$, with $\vartheta\in [0,2\pi)$, by inserting \eqref{eq:rho0} into \eqref{eq:Heringopp} we obtain:
\beq
\rho(1,\vartheta) = \frac{1+2\cos \vartheta}{4} \rho(1,0)+\frac{1-2\cos \vartheta}{4} \rho(1,\pi)+\frac{1+2\sin \vartheta}{4} \rho\left(1,\frac{\pi}{2}\right)+\frac{1-2\sin \vartheta}{4} \rho\left(1,\frac{3\pi}{2}\right),
\eeq 
which shows that each pure hue can be seen as a the pure state obtained by superposition (in the quantum sense) of the four pure states identified with red, green, yellow and blue.

The compatibility between the trichromatic theory and Hering's one (physiologically supported by the action of the three retinal photoreceptors and that of ganglion cells, respectively) is inherent in the mathematical framework that we are discussing. This result is far from obvious: to the best of our knowledge, the only other way in which the two theories can be mathematically re-conciliated is via an \textit{a-posteriori} principal component analysis on a database of color images in the framework of natural image statistics, see e.g. \cite{Buchsbaum:83,Ruderman:98,Provenzi:16}. 

The \textit{expectation value} of an observable $a\in \mathcal A$ on a state described by a density matrix $\rho_ {\bf s}$ represents the average outcome after multiple measurements of the observable when the system is prepared each time in the same state. Mathematically, it can be computed as follows:
\beq
\langle a \rangle_{\rho_ {\bf s} }= \Tr(\rho_{  {\bf s}} a).
\eeq 
The computation of the expectation value of the two real Pauli matrices $\sigma_1$ and $\sigma_2$ gives:
\beq \langle \sigma_1 \rangle_{\rho(r,\vartheta)} =\text{Tr}(\rho(r,\vartheta)  \sigma_1)=r\cos \vartheta, \quad \langle \sigma_2 \rangle_{\rho(r,\vartheta)} =\text{Tr}(\rho(r,\vartheta)  \sigma_2)=r\sin \vartheta,
\eeq 
which, compared to \eqref{eq:Heringopp},  shows that the expectation values of the real Pauli matrices $\sigma_1$ and $\sigma_2$ give the information about the \textit{degree of opposition} R-G and Y-B, respectively.

In \cite{Berthier:2021JofImaging}, a fundamental theoretical result that corroborates the hypothesis about the quantum nature of color perception has been obtained: even if, as shown before, the two degrees of chromatic opposition are intrinsic characteristics of a chromatic state, their simultaneous measurement may be subject to uncertainty. To state this fact precisely, we recall that the quadratic dispersion (or variance) of an observable $a$ on the state $\rho_ {\bf s}$ is:
\beq
(\Delta a_{\rho_ {\bf s}})^2:=\langle a^2 \rangle_{\rho_ {\bf s}} - (\langle a \rangle_{\rho_ {\bf s}})^2 = \text{Tr}(\rho_ {\bf s} a^2)-(\text{Tr}(\rho_ {\bf s}  a))^2. 
\eeq 
If we consider the observables given by the two real Pauli matrices, then we get:
\beq\label{eq:SchrPauli}
(\Delta {\sigma_1}_{\rho_ {\bf s}})^2(\Delta {\sigma_2}_{\rho_ {\bf s}})^2 \geq  \frac{r^4}{4} \sin^2(2\vartheta),
\eeq 
for all $r\in [0,1]$ and $\vartheta \in [0,2\pi)$. The interpretation of formula (\ref{eq:SchrPauli}) is the following: no uncertainty relations are present for the achromatic state or for chromatic states lying on the axes R-G and Y-B, however, for $r\in (0,1]$ and $\vartheta \in [0,2\pi)\setminus \{0,\frac{\pi}{2},\pi,\frac{3}{2}\pi\}$, there exists a lower bound strictly greater than 0 for the product of quadratic dispersions of $\sigma_1$ and $\sigma_2$ on the state defined by $\rho(r,\vartheta)$. This lower bound is a non-linear function of the variables $(r,\vartheta)$ and it is maximum for pure hues, $r=1$, halfway in between G and B, B and R, R and Y and Y and G, i.e. $\vartheta = \frac{\pi}{4}+k\frac{\pi}{2}$, $k=1,2,3$. 

The quantum-like system that we have described is well-known in quantum information, where it is called a \textit{rebit}, i.e. a real qubit \cite{Wootters:2014}. A (complex) qubit is a quantum system that can only be \textit{measured} in two states, in clear analogy with the classical concept of bit. In the quantum-like framework of color perception, the two rebit states are the degrees of chromatic opponency, for this reason such a rebit has been named `Hering's rebit'  in \cite{Berthier:2020}. This latter can be thought as a mathematical formalization of the celebrated Newton chromatic disk \cite{Newton:52}.


\section{From perceptual to perceived colors through quantum measurements}\label{sec:quantumeasurement}

In this section,  the paramount important concept of color measurement is introduced in a systematic way into our analysis from the perspective of the well-established theory of quantum measurement, see e.g.  \cite{Moretti:17,Busch:97,Heinosaari:2011,Busch:16}.

Before dealing with technical details, we deem important to introduce the concept of color  measurement by honoring Maxwell's brilliant idea about color matching, which has been the only way of measuring colors for more than a century: this technique consists in fixing a test color stimulus on one side of a bipartite field and asking to a human observer to construct another color stimulus, called match, by superimposing three colored lights so that the test and match stimuli are perceived as the same color when observed  side-by-side.

Maxwell's color measurement method still remains the most widely used, even if more advanced techniques have been developed, thanks in particular to the possibilities offered by the use of computer monitors in psycho-visual experiments. 

Crucially, the result of identically prepared psycho-visual tests in general do not correspond to a precise selection of the match, but to a distribution of choices picked around the most probable one. As we are going to discuss in detail in the following subsections, also the theory of quantum measurement is inherently related to a probabilistic interpretation of the outcome of an experiment, thus providing a strong motivation for the pertinence of our approach.

\subsection{POVM, unsharp observables and effects}\label{subsec:POVMandeffects}
In 1927, von Neumann established with the two papers \cite{vonNeumann:271,vonNeumann:272} the association between a quantum observable,  represented by a Hermitian operator $A$ on the  Hilbert state space $(\mathcal H, \langle \; , \; \rangle)$ of the system, and the unique spectral measure $E_A$ related to $A$ by the spectral resolution theorem. $E_A$ is a projector-valued measure (PVM) and its properties lead to the following fundamental result: if $\rho$ is a density matrix representing a state and $\mathscr B(\R)$ is the Borel $\sigma$-algebra of the real line, then the map: 
\beq
\mathscr B(\R)\ni B\mapsto p^{E_A}_\rho(B):=\Tr(\rho E_A(B))\in [0,1],
\eeq  
is a probability measure on $\R$, see e.g. \cite{Moretti:17}. The so-called \textit{minimal  interpretation}, see \cite{Busch:97}, establishes that the number $p_\rho^{E_A}(B)$ represents the probability that, when the system is prepared in the state $\rho$, the measurements of $A$ is registered in the Borel set $B$.

The influence of von Neumann on the foundation of quantum mechanics was so strong that only in the late 1960's Ludwig, Davies and Lewis, see \cite{Ludwig:64,Davies:70}, realized that requiring $E$ to be a PVM is excessive if one considers the probability $p_\rho^{E}(B)$ the only meaningful information about the outcome of a measurement of the observable $A$ on the state $\rho$.

In fact, the request can be relaxed by asking $E$ to be just a  semi-spectral measure, more commonly known as a Positive Operator-Valued Measure (POVM) on a generic measurable space $(\Omega,\mathcal F)$, where  $\Omega$ is called \textit{value space}  or \textit{outcome space}, and $\mathcal F$ is a $\sigma$-algebra. This amounts to requiring that, for all measurable set $X\in \mathcal F$, $E(X)$ is a positive operator bounded between the null and the identity operators, $\textbf{0}$ and $id_{\mathcal H}$ respectively, on $\mathcal H$ and not necessarily a projector, this latter representing just a special case for a POVM. 

Of course, the condition $\textbf{0}\le E(X) \le id_{\mathcal H}$ must be interpreted with respect to the partial ordering of positive operators on $\mathcal H$, i.e. $\langle x,E(X)x\rangle \ge 0$ and $\langle x,(id_{\mathcal H}-E(X))x\rangle \ge 0$ for all $x\in \mathcal H$ and all $\mathcal F$-measurable set $X$. The following definition has been introduced by Ludwig \cite{Ludwig:64} and popularized by Kraus \cite{Kraus:83}.

\begin{definition}
	For all $X\in \Omega$, the positive operator $E(X)$ in the range of a \emph{POVM} is called an effect on $\mathcal H$. The set of all effects on $\mathcal H$ is denoted with $\mathcal E(\mathcal H)$ and called effect space of $\mathcal H$.
\end{definition} 
It is easy to verify that $\mathcal E(\mathcal H)$ is a convex partially ordered subset of the cone of positive operators on $\mathcal H$. The partial ordering of $\mathcal E(\mathcal H)$ is connected with the  probability ordering in the following sense: given $E_1,E_2\in \mathcal E(\mathcal H)$, we have $E_1\le E_2$ if and only if $\Tr(\rho E_1)\le \Tr(\rho E_2)$ for all state density matrix $\rho$. Instead, the convexity of $\mathcal E(\mathcal H)$ reflects the possibility to combine  measurements to create new ones. 

\begin{definition} 
	An unsharp (or generalized) observable is defined by either a \emph{POVM} or the collection of effects $\{E(X)\}_{X\in \mathcal F}$ such that $E: \Omega \to \mathcal E(\mathcal H)$, is a \emph{POVM}. Conversely, an observable defined by a PVM is called sharp.
\end{definition} 

The notion of unsharpness is crucial for our color perception model: this concept is intended to describe \textit{intrinsic limitations} in the possibility to prepare states or perform measurements with absolute precision, as, on the contrary, it would be the case for a sharp observable. 

An illuminating example of unsharp observable is provided by Busch et al. in \cite{Busch:97}: a more faithful `laboratory report' of the celebrated Stern-Gerlach experiment for spin measurement reveals that what is commonly written in quantum mechanical textbooks is actually an idealization. In the real experiment, the spin-$1/2$ particles that passed through the inhomogeneous magnetic field were indeed split in two beams before impinging on a glass plate, however the two beams were \textit{distinguishable but not disjoint}. When this important experimental detail is taken into account, the outcome of the Stern-Gerlach experiment is more correctly interpreted as the manifestation of an unsharp spin observable, rather than a sharp one.

\subsection{From the infinite cone of perceptual colors to the finite color solid of perceived colors}\label{subsec:colorsolid}

The color perception described by Schrödinger's axioms is idealized because of the so-called first axiom, due to Newton \cite{Newton:52}, which says that if $c$ is a perceptual color, than also $\lambda c$ is so, for \textit{all} $\lambda \in \R^+$. 
Already Resnikoff pointed out in \cite{Resnikoff:74}  the idealized nature of this axiom and the lack of correspondence with the so-called \textit{glare limit}, i.e. the fact that light  stimuli with an intensity larger than a suitable value, that depends on the observational conditions, cannot be perceived since they cause a complete saturation in the photoreceptors response, see e.g. \cite{Cie:95,Provenzi:16modernphysics,Provenzi:20}. 

Since the trichromacy axiom is a consequence of Schrödinger and Resnikoff's axioms, the perceptual color space $\mathcal C$, as we have seen in section \ref{sec:recap}, is represented by an infinite cone: either positive-definite $2\times 2$ real matrices or the future lightcone in the 3-dimensional Minkowski space. 

This is the reason why we have chosen to use the term \textit{perceptual} colors and not \textit{perceived} colors in the statement of the trichromacy axiom: perceptual colors are associated to \textit{ideal} observers, able to measure any kind of light stimuli, while perceived colors are associated to \textit{real} observer, that can perform meaningful measurements only on a bounded set of light stimuli, producing a set of outcomes confined in a finite volume. 

One of the first scientists to grasp this concept  was the chemist W. Ostwald in his celebrated `color primer'  theory, see e.g. \cite{Granville:94}, in which the space of perceived colors is represented by a finite convex double cone commonly called `color solid'. More recent contributions in this sense have been made by deValois et al. \cite{deValois:97,deValois:2000} and Koenderink et al. \cite{Koenderink:03, Koenderink:10}.

In what follows, we are going to show, as first argued in section 5 of  \cite{Berthier:2020}, that effects provide an elegant and natural way to pass from the infinite cone of perceptual colors to the finite double cone of perceived colors.

First of all, we fix the measurable space $(\Omega,\mathcal F)$ to define POVMs in the case of color perception. For trichromatic observers, the outcome space $\Omega$ is $\R^3$ and, at least for the moment, there is no reason to consider a $\sigma$-algebra different than the Borel one, i.e. $\mathscr B(\R^3)$. 

Then, we re-write $\overline{\mathcal C}(\mathcal A)$, $\mathcal A=\hr,\spinf$, to make states appear explicitly as follows:
\beq\label{eq:twoalpharho}
\overline{\mathcal C}(\mathcal H(2,\R)) = \left\{2\alpha \rho_{\bf s}=\begin{pmatrix}
	\alpha (1+s_1) & \alpha s_2 \\ \alpha s_2 & \alpha (1-s_1)
\end{pmatrix},\; \alpha\geq 0, \; {\bf v_s}=(s_1,s_2)\in\mathcal{D}\right\} \cong \overline{\mathcal H^+}(2,\R)
\eeq
and
\beq\label{eq:twoalphasv}
\overline{\mathcal C}(\R\oplus \R^2) = \left\{2\alpha\chi(\rho_{\bf s})=\begin{pmatrix}
	\alpha\\  \alpha {\bf v_s}
\end{pmatrix},\; \alpha\geq 0, \; {\bf v_s}\in\mathcal{D}\right\} \cong \overline{\mathcal L^+}.
\eeq 
Effects form a closed convex subset $\mathcal E$ of $\overline{\mathcal C}(\mathcal A)$, that will be called the \textit{effect space}. The easiest way to characterize effects is by considering a matrix $\eta_{\bf e}\in \overline{\mathcal C}(\mathcal H(2,\R))$ bounded between the null and the identity $2\times 2$ matrix or, equivalently, by the element $\chi(\eta_{\bf e})\in \overline{\mathcal C}(\R\oplus \R^2)\cong \overline{\mathcal L^+}$.  

It is useful to adopt a general symbol to denote an effect $\bf e$ when it is not important to know if it is realized as the matrix $\eta_{\bf e}$ or the vector $\chi(\eta_{\bf e})$. We will use the following notation:
\beq
{\bf e}:=(e_0,{\bf v_e}),
\eeq 
where $e_0$ is the \textit{effect magnitude} and $\bf v_e$ is the \textit{effect vector}, defined as follows
\beq
{\bf v}_{\bf e}:=\left(\frac{e_1}{e_0},\frac{e_2}{e_0}\right),
\eeq 
$e_0,e_1,e_2\in \R$. In this way we have:
\begin{equation}\label{eq:mateff}
	\eta_{\bf e}=\left(\begin{array}{cc}e_0+e_1 & e_2 \\e_2 & e_0-e_1\end{array}\right) \in \hr,
\end{equation}
and  
\begin{equation}\label{eq:chietae}
	\chi(\eta_{\bf e})=e_0 \begin{pmatrix}
		1 \\ \textbf{v}_ {\bf e}
	\end{pmatrix}\in \R\oplus \R^2.
\end{equation}
If ${\bf v_e}=\bf 0$, we write ${\bf e}=\bf e_a$ and call it an \textit{achromatic effect}. It is clear that $\eta_{\bf e_a}=e_0\sigma_0$.

\begin{figure}[htbp]
	\begin{center}
		\includegraphics[width=5cm]{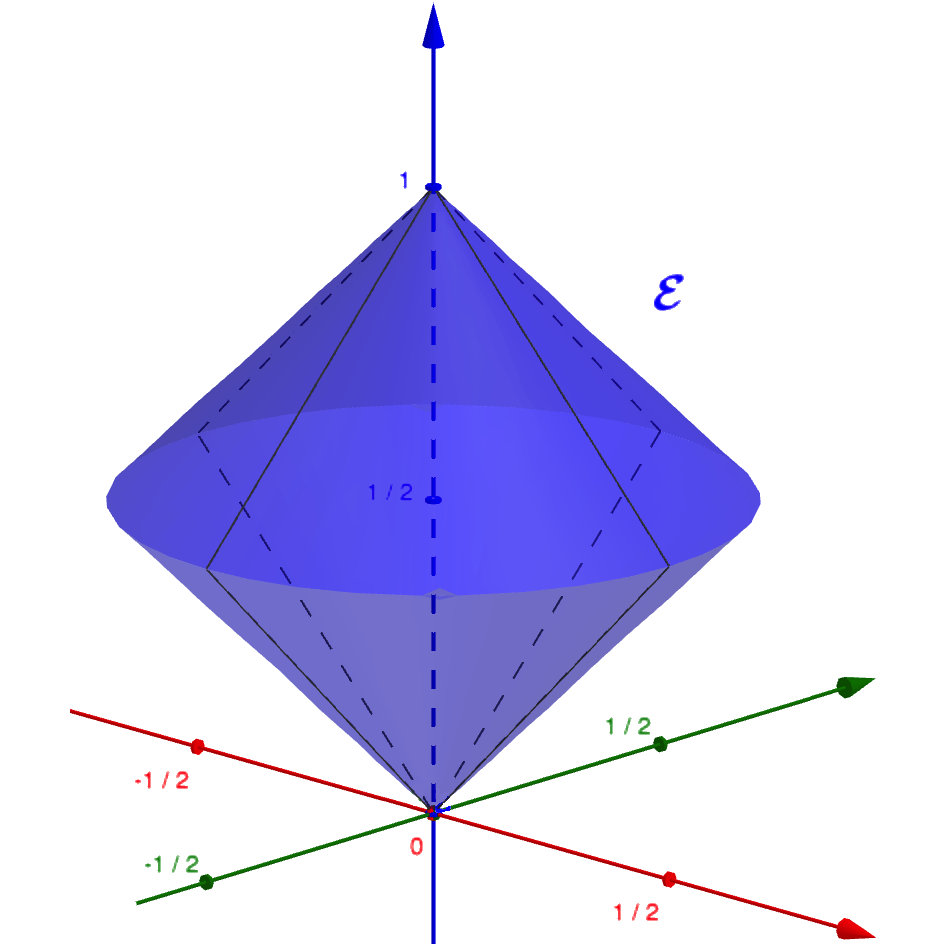}
		\caption{The convex double cone representing the effect space.}
		\label{fig:doubleconeeffects}
	\end{center}
\end{figure}

$\eta_{\bf e}$ defines an effect if and only if ${\bf 0}\le \eta_{\bf e}\le \sigma_0$, this double inequality is equivalent to the request that the determinant and the trace of both $\eta_{\bf e}$ and $\sigma_0-\eta_{\bf e}$ are non-negative. From $\det(\eta_{\bf e})\ge 0$ we obtain ${\bf v}_{\bf e}\in \mathcal{D}$ and, by considering all the other constraints, we find that the effect space can be geometrically characterized in an explicit way as follows:
\begin{equation}\label{eq:effect1}
	\mathcal E=\left\{(e_0,e_1,e_2)\in \R^3, \; e_0 \in [0,1], \;  e_1^2+e_2^2\le \min\limits_{e_0 \in [0,1]} \left\{(1-e_0)^2,e_0^2\right\} \right\}.
\end{equation}
$\mathcal E$ is a closed convex double cone with a circular basis of radius $1/2$ located height $\alpha=1/2$ and with vertices in $(e_0,e_1,e_2)=(0,0,0)$ and $(e_0,e_1,e_2)=(1,0,0)$. 
The graphical representation of $\mathcal E$ is depicted in Figure \ref{fig:doubleconeeffects} and it is in perfect agreement with the geometry of perceived colors advocated by Ostwald and deValois.

We can obtain another equivalent expression $\mathcal E^*$ of the effect space by considering the dual cone $\overline{\mathcal C}^*( \mathcal A)$, thanks to \eqref{eq:dualcone} we have that:
\beq
\mathcal E^*=\left\{ (e_0, e_1, e_2) \in (\R^3)^* \; : \;  (e_0, e_1, e_2) \begin{pmatrix}
	1 \\ {\bf v_s}
\end{pmatrix}=e_0+e_1s_1+e_2s_2\in [0,1], \;  {\bf v_s}\in \mathcal D \right\}.
\eeq 
This alternative expression shows that an effect can also be identified with the affine function acting on the set of states given by ${\bf e}: {\cal S}(\mathcal A)\to [0,1] $, ${\bf e}({\bf s})=e_0+e_1s_1+e_2s_2$, which represents the probability of the outcome $(e_0,e_1,e_2)$ after a measurement on the visual scene prepared in the state $\bf s$.

Now, in view of the conceptual meaning of unsharp observables and of the probabilistic nature of color measurement recalled at the beginning of section \ref{sec:quantumeasurement}, it seems that the most pertinent description of a perceived color is that of a sensation generated by the measurement of an unsharp observable, which, translated in mathematical terms, leads to the following fundamental definition.

\begin{definition}
	Let $\mathcal A$ be either $\hr$ or $\spinf$. The  act of measuring a perceptual color belonging to $\overline{\mathcal C}(\mathcal A)$ is identified  with an unsharp observable represented by a $\mathcal E$-valued  \emph{POVM}. A perceived color is the outcome of the measurement, i.e. an effect belonging to $\mathcal E$. 
\end{definition} 

In summary, our quantum-like description of color measurement is represented by the couple $(\rho_ {\bf s},E)$, where $\rho_ {\bf s}$ is a density matrix representing the state $\bf s$ in which a visual scene has been prepared for  measurements of perceptual colors and $E$ is a $\mathcal E$-valued POVM. $\rho_ {\bf s}$ and $E$ are in \textit{statistical duality} in the sense that the number $\Tr(\rho_ {\bf s} E(X))$ represents the probability of registering the measurement outcome in $X\in \mathscr B(\R^3)$, where $E(X)$ is an effect belonging to the range of $E$. As a consequence of the statistical duality state-effects, one can reverse their role and consider the action of states of the set of effects as follows: every state $\bf s$ defines a function $f_{\bf s}:\mathcal E(\mathcal S(\mathcal A))\to [0,1]$ such that, for all ${\bf e}\in \mathcal E(\mathcal S(\mathcal A))$, $f_{\bf s}({\bf e}):={\bf e(s)}$.


\section{The relationship between quantum color measurements and the relativistic nature of color perception}\label{sec:colormeasurelativistic}
In this section we explore the profound and somewhat surprising relationship between quantum color measurement and the relativistic nature of color perception, establishing results that extend and generalize those obtained in the paper \cite{Berthier:21JMP}.

\subsection{Lüders operations, Kraus operators  and state transformations after nondestructive measurements}\label{subsec:Luders}
In this subsection we recall a crucial concept from quantum measurement theory: how to describe a change of state of a quantum system induced by a measurement. This will turn out to be the key to understand how relativistic transformations appear in our quantum-like theory of color perception. Dealing with the general theory would introduce an unnecessary degree of mathematical sophistication, thus we chose to adapt directly the general results to our needs. The main references are the books \cite{Kraus:83,Busch:97,Busch:16}.

As previously discussed, we associate the process of color perception by a human observer with a quantum measurement on a given state of a visual scene. Since the act of observing does not disrupt a visual scene, color perception can be more precisely associated to a \textit{nondestructive measurement}. This remark makes a quantum-like theory of color perception an interesting ground to test the quantum measurement formalism, since the majority of interactions used to perform quantum measurements in the physical domain are destructive. 

An important function associated to this kind of measurements in quantum theory is called operation, a concept first introduced by Haag and Kastler in 1964, see \cite{Haag:64}, as an essential element of their axiomatic algebraic quantum field theory.

\begin{definition}
	An operation is a convex-linear positive function  $\phi:\overline{\mathcal C}(\mathcal A)\to \overline{\mathcal C}(\mathcal A)$ satisfying the constraint $0\le \Tr(\phi(a))\le 1$ for all $a\in \overline{\mathcal C}(\mathcal A)$.
\end{definition}  
Here we are actually interested in the restriction of $\phi$ to the subset $\mathcal S(\mathcal A)$ of $\overline{\mathcal C}(\mathcal A)$ given by the density matrices $\rho_ {\bf s}\in \mathcal S(\hr)$, or, equivalently, $\chi(\rho_ {\bf s}) \in \mathcal S(\spinf)$. For simplicity we will also denote this restriction with $\phi$. The request that $\phi$ acts linearly on convex combinations is natural if one wants to be consistent with the mixing property of states. Instead, it may seem odd to demand  $\Tr(\phi(\rho_ {\bf s}))\in[0,1]$, because  $\mathcal S(\mathcal A) \subsetneq \phi(\mathcal S(\mathcal A))=:\tilde{\mathcal S}(\mathcal A)$, i.e. $\rho_ {\bf s}\in \mathcal S(\mathcal A)$ will lose the property of having unit trace after an operation, becoming a so-called \textit{generalized density matrix} $\phi(\rho_ {\bf s})\in \tilde{\mathcal S}(\mathcal A)$.  

Notice that this is analogous to what happens when passing from a PVM to a POVM, where the norm of the operator in the range of the measure passes from the fixed value of 1 to a number between 0 and 1. In this case it  is the trace of the density matrix that undergoes the same relaxation process which, as observed by Kraus \cite{Kraus:83}, is advantageous because this trace carries useful additional information of probabilistic nature. 

To formalize this statement, let us consider particular state transformations, called \textit{Lüders operations}, related to a change of state after a measurement associated to an effect.  If the original state ${\bf s}$ of a visual scene is described by the density matrix $\rho_ {\bf s}$, then the measurement of a perceived color, identified with the effect $  {\bf e}$ and represented by the matrix $\eta_ {\bf e}$, provokes a change of the initial state $\bf s$. In quantum measurement theory, the \textit{Lüders operation associated to the effect} ${\bf e}$ is the correspondence 

\begin{equation}\label{eq:psies}
	\begin{array}{cccl}
		\psi_{\bf e}: &  \mathcal S(\mathcal A) & \longrightarrow & \tilde{\mathcal S}(\mathcal A)  \\
		& {\bf s} & \longmapsto & \psi_{\bf e}({\bf s}):=\eta_ {\bf e}^{1/2} \rho_ {\bf s} \eta_ {\bf e}^{1/2},
	\end{array}
\end{equation}
$\eta_ {\bf e}^{1/2}$ is called the \textit{Kraus operator} associated to $\psi_ {\bf e}$ and it is the square root of $\eta_ {\bf e}$, i.e. the only positive semi-definite matrix such that $\eta_ {\bf e}^{1/2}\eta_ {\bf e}^{1/2}=\eta_ {\bf e}$.

Thanks to the cyclic property of the trace we have:
\beq
\Tr(\psi_ {\bf e}(  {\bf s}))= \Tr(\eta_ {\bf e}^{1/2} \rho_ {\bf s} \eta_ {\bf e}^{1/2}) = \Tr(\rho_ {\bf s} \eta_ {\bf e}) = \langle   {\bf e} \rangle_ {\bf s}.
\eeq
This implies that 
\beq\label{eq:phies}
\varphi_ {\bf e}(  {\bf s})=\frac{\psi_ {\bf e}(  {\bf s})}{\langle   {\bf e} \rangle_ {\bf s}},
\eeq 
is a density matrix representing the post-measurement chromatic state, which appears fused together with the probabilistic information carried by $\langle   {\bf e} \rangle_ {\bf s}$ in the generalized density matrix $\psi_ {\bf e}(  {\bf s})$.

By direct computation it can be verified that the expectation value of $\bf e$ on $\bf s$ can be written more explicitly as follows:
\beq\label{eq:expes}
\langle {\bf e} \rangle_{\bf s} = \Tr(\rho_{\bf s} \eta_{\bf e}) = e_0(1+{\bf v_e} \cdot {\bf v_s}) = 2 \chi(\rho_{\bf s}) \cdot \chi(\eta_{\bf e}),
\eeq 
where $\cdot$ represents the Euclidean inner product.

\subsection{Quantum measurements and Lorentz boosts}\label{subsec:boosts}

The aim of this section is to show that, given an effect $\bf e$, the Lüders operation $\psi_{\bf e}:\mathcal S(\R\oplus \R^2)\to \mathcal {\tilde S}(\R\oplus \R^2)$ representing the state transformation of a visual scene induced by the measurement of the perceived color $\bf e$  can be explicitly represented by a convex-linear transformation proportional to a Lorentz boost on $\R^{1,2}$. Analogous results to the ones that we are going to present have been proven in \cite{Arrighi:04} in the context of theoretical physics. Here, however, we avoid using tensor computations by adopting an alternative methodology which leads to a simpler proof.

By \eqref{eq:psies}, the only information that we need in order to find $\psi_ {\bf e}$ is the explicit expression of $\eta_ {\bf e}^{1/2}$, this will already give us a hint of the fact that the state change $\psi_{\bf e}$ is related to a relativistic transformation because of the natural appearance of the Lorentz factor. The second, trickier, step of our procedure consists in observing that the expression of $\chi(\psi_ {\bf e}(  {\bf s}))\in \tilde{\mathcal S}(\spinf)$ is related to the spin representation of density matrices via the key eq. \eqref{eq:chipsi}, this  turns out to be the hidden link with Lorentz boosts. 

So, let us first concentrate on the computation of the Kraus operator $\eta_ {\bf e}^{1/2}$. To this aim, let us re-write the matrix $\eta_{\bf e}$ in eq. (\ref{eq:mateff}) as follows $\eta_{\bf e}=2e_0\rho_{  {\bf e}/2e_0}$, where 
\begin{equation}
	\rho_{{\bf e}/2e_0}={1\over 2}\left(\begin{array}{cc}1+e_1/e_0 & e_2/e_0 \\e_2/e_0 & 1-e_1/e_0\end{array}\right)
\end{equation}
is a density matrix because  $\textbf{v}_ {\bf e}=(e_1/e_0,e_2/e_0))\in \mathcal D$. The square root of $\eta_{\bf e}$ is simply:
\begin{equation}\label{eq:etarho}
	\eta_{\bf e}^{1/2}=\sqrt{2e_0}\rho_{{\bf e}/2e_0}^{1/2}.
\end{equation}
This shows that, to be able to compute $\eta_ {\bf e}^{1/2}$, it is sufficient to know how to compute the square root of a density matrix $\rho_ {\bf s}$. Since in our theory such matrices are $2\times 2$, we have \cite{Levinger:80}:
\beq
\rho_ {\bf s}^{1/2}=\frac{1}{A}(B\sigma_0 + \rho_ {\bf s}),
\eeq 
where 
\beq
B=\sqrt{\det(\rho_ {\bf s})}=\frac{1}{2}\sqrt{1-\|\textbf{v}_ {\bf s}\|^2}  \quad \text{ and } \quad    A=\sqrt{\Tr(\rho_ {\bf s})+2B}=\sqrt{1+\sqrt{1-\|\textbf{v}_ {\bf s}\|^2}}.
\eeq 
We can distinguish two cases:
\begin{enumerate}
	\item if $\|\textbf{v}_ {\bf s}\|=1$, i.e. if  $\rho_ {\bf s}$ is a pure state, then $B=0$, so $A=\sqrt{\Tr(\rho_{\bf s})}=1$, hence $\rho_ {\bf s}^{1/2}=\rho_ {\bf s}$;
	\item otherwise, when $0\le \|\textbf{v}_ {\bf s}\|<1$, we can define a Lorentz-like factor as follows:
	\begin{equation}
		\gamma_{\bf s}:={1\over \sqrt{1-\Vert {\bf v}_{\bf s}\Vert^2}},
	\end{equation}
	by re-writing $1/A$ and $B$ in terms of $\gamma_{\bf s}$, we arrive at the following expression of the square root $\rho_ {\bf s}$ in which the Lorentz-like factor $\gamma_ {\bf s}$ appears explicitly:
	\begin{equation}\label{eq:rhosquare}
		\rho_{\bf s}^{1/2}={\sqrt{\gamma_{\bf s}}\over\sqrt{1+\gamma_{\bf s}}}\left({1\over 2\gamma_{\bf s}}\sigma_0+\rho_{\bf s}\right).
	\end{equation}
\end{enumerate} 

Let us now pass to the second step of our procedure. If we interpret the map $\chi$ defined in \eqref{eq:isoH2} as a linear isomorphism between $\hr$ and the 3-dimensional Minkowski space $\R^{1,2}$, then we can consider the spin representation of the vector space of $2\times 2$ real matrices $\mathcal M(2,\mathbb R)$ into $\text{End} (\mathbb R^{1,2})$, the vector space of  endomorphisms of the Minkowski space, defined as follows:
\begin{displaymath}
	\begin{array}{cccl}
		\xi : & \mathcal M(2,\mathbb R)  & \longrightarrow &    \text{End} (\mathbb R^{1,2})\\
		& A & \longmapsto         & \xi(A):=\chi\circ {\rm Ad}_A\circ\chi^{-1},
	\end{array}
\end{displaymath}
such that, for all $\omega \in \hr$, 
\begin{equation}\label{eq:chiomegaA}
	\xi(A)(\chi(\omega))=\chi(A\omega A^t).
\end{equation}
Combining eq. \eqref{eq:chiomegaA} with eqs.  \eqref{eq:psies}, (\ref{eq:etarho}) and considering the symmetry of $\eta_ {\bf e}^{1/2}$ we find:
\beq
\chi(\psi_{\bf e}({\bf s}))=\chi(\eta_ {\bf e}^{1/2} \rho_ {\bf s} (\eta_ {\bf e}^{1/2})^t)=\xi(\eta_ {\bf e}^{1/2}) (\chi(\rho_ {\bf s}))=\xi(\sqrt{2e_0}\rho_{{\bf e}/2e_0}^{1/2}) (\chi(\rho_ {\bf s}))
\eeq 
i.e.
\begin{equation}\label{eq:chipsi}
	\chi(\psi_{\bf e}({\bf s}))=2e_0\xi(\rho_{{\bf e}/2e_0}^{1/2})(\chi(\rho_{\bf s})).
\end{equation}
\begin{proposition}
	Let ${\bf s}$ be a state of Hering's rebit parameterized by the vector $\emph{\textbf{v}}_ {\bf s}=(s_1,s_2)^t\in \mathcal D$ and let $\rho_{\bf s}$ be the corresponding density matrix. Then, the following assertions hold.
	
	\begin{enumerate}
		\item If $\rho_ {\bf s}$ represents a pure state, i.e. $\|\emph{\textbf{v}}_{\bf s}\|=1$, then the matrix representation of $\xi(\rho_{\bf s}^{1/2})$ with respect to the canonical basis of $\R^{1,2}$ is:
		\begin{equation}\label{eq:xipure}
			[\xi(\rho_{\bf s}^{1/2})]= \frac{1}{2} \begin{pmatrix}
				1 & s_1 & s_2 \\
				s_1 & s_1^2 & s_1s_2\\
				s_2 & s_1s_2 & s_2^2
			\end{pmatrix}\equiv {1\over 2} \begin{pmatrix} 1 & {\bf v}_{\bf s}^t \\
				{\bf v}_{\bf s} & {\bf v}_{\bf s}{\bf v}_{\bf s}^t
			\end{pmatrix}.
		\end{equation}
		\item If $\rho_ {\bf s}$ represents a mixed state, i.e. $\|\emph{\textbf{v}}_{\bf s}\|<1$, then the spin representation of  $\rho_ {\bf s}^{1/2}$ is the endomorphism of the 3-dimensional Minkowski space given by:
		\begin{equation}\label{prop1}
			\xi(\rho_{\bf s}^{1/2})={1\over 2\gamma_{\bf s}}B({\bf v}_{\bf s}),
		\end{equation}
		where $B({\bf v}_{\bf s})$ is the Lorentz boost parametrized by ${\bf v}_{\bf s}$. The matrix representation of $B({\bf v}_{\bf s})$ with respect to the canonical basis of $\R^{1,2}$ is:
		\begin{equation}\label{eq:xinonpure}
			\left[B({\bf v}_{\bf s})\right]=\begin{pmatrix}
				\gamma_{\bf s} & \gamma_{\bf s} s_1 & \gamma_{\bf s} s_2\\
				\gamma_{\bf s} s_1 & 1+\frac{\gamma_ {\bf s}^2}{1+\gamma_ {\bf s}}s_1^2 & \frac{\gamma_ {\bf s}^2}{1+\gamma_ {\bf s}}s_1s_2  \\
				\gamma_{\bf s}  s_2 & \frac{\gamma_ {\bf s}^2}{1+\gamma_ {\bf s}}s_1s_2  & 1+\frac{\gamma_ {\bf s}^2}{1+\gamma_ {\bf s}}s_2^2 
			\end{pmatrix} \equiv \begin{pmatrix}
				\gamma_{\bf s} & \gamma_{\bf s}{\bf v}_{\bf s}^t \\\gamma_{\bf s}{\bf v}_{\bf s} & \sigma_0 +{\gamma_{\bf s}^2\over 1+\gamma_{\bf s}}{\bf v}_{\bf s}{\bf v}_{\bf s}^t
			\end{pmatrix}.
		\end{equation}
	\end{enumerate} 
\end{proposition}

\proof The proof consists in exhibiting explicitly the matrix entries. We start by showing  computations that are common to both cases and then we specialize our reasoning distinguishing the purity or not of the state $  {\bf s}$.

Let $\rho_ {\bf s}$ be a density matrix and let $\left[\xi(\rho_{\bf s}^{1/2})\right]$ denote the matrix of the endomorphism $\xi(\rho_{\bf s}^{1/2})$ with respect to the canonical basis $(e_k)_{k=0,1,2}$ of $\R^{1,2}$, then, by recalling that $\sigma_k=\chi^{-1}(e_k)$ for all $k=0,1,2$, we have:
\beq
\begin{split}
	\left[\xi(\rho_{\bf s}^{1/2})\right]_{ij} & =e_i^t \,  \xi(\rho_{\bf s}^{1/2}) e_j = e_i^t \, (\chi \circ  \text{Ad}_{\rho_{\bf s}^{1/2}} \circ \chi^{-1})(e_j) =  e_i^t \, (\chi \circ \text{Ad}_{\rho_{\bf s}^{1/2}}) (\sigma_i) \\
	& = e_i^t \, \chi (\rho_{\bf s}^{1/2} \sigma_i \rho_{\bf s}^{1/2}) = \chi (\rho_{\bf s}^{1/2} \sigma_i \rho_{\bf s}^{1/2}) \cdot \chi(\sigma_i),
\end{split}
\eeq 
for $i,j=0,1,2$, so, thanks to eq. \eqref{eq:expes}, 

\begin{equation}\label{eq:matentries}
	\left[\xi(\rho_{\bf s}^{1/2})\right]_{ij}=\chi(\rho_{\bf s}^{1/2}\sigma_j\rho_{\bf s}^{1/2})\cdot\chi(\sigma_i)={1\over 2}{\rm Tr}(\rho_{\bf s}^{1/2}\sigma_j\rho_{\bf s}^{1/2}\sigma_i), \quad i,j=0,1,2.
\end{equation}
This matrix is symmetric thanks to the cyclic property of the trace. Moreover, we note that 
\begin{equation}
	\left[\xi(\rho_{\bf s}^{1/2})\right]_{j0}=\left[\xi(\rho_{\bf s}^{1/2})\right]_{0j}={1\over 2}{\rm Tr}(\rho_{\bf s}\sigma_j)={s_j\over 2},
\end{equation}
for $j=0,1,2$, with $s_0=1$. Hence, in the next computations we only need to consider $i,j=1,2$.

\begin{enumerate}
	\item we have seen that when $\rho_ {\bf s}$ represents a pure state we have $\rho_ {\bf s}^{1/2}=\rho_ {\bf s}=\rho_ {\bf s}^2$ and so eq.  \eqref{eq:matentries} can be re-written as:
	\beq
	\left[\xi(\rho_{\bf s}^{1/2})\right]_{ij}={1\over 2}{\rm Tr}(\rho_{\bf s}\sigma_j\rho_{\bf s}\sigma_i), 
	\eeq 
	one can verify the validity of eq. \eqref{eq:xipure} by computing directly the trace in the equation above and keeping in mind that, in the present case, $s_1^2+s_2^2=1$.
	
	\item In this case we have $\|\textbf{v}_ {\bf s}\|<1$, so $\rho_ {\bf s}$ represents a mixed state and $\gamma_ {\bf s}$ is well-defined. Let us consider $i=j=1$, then, thanks to eq.  \eqref{eq:rhosquare} and to the fact that $\sigma_1^2=\sigma_2^2=\sigma_0$ we have 
	\begin{equation}
		\begin{split}
			\rho_{\bf s}^{1/2}\sigma_1\rho_{\bf s}^{1/2}\sigma_1 & ={\gamma_{\bf s}\over1+\gamma_{\bf s}}\left({1\over 2\gamma_{\bf s}}\sigma_0+\rho_{\bf s}\right)\sigma_1 \left({1\over 2\gamma_{\bf s}}\sigma_0+\rho_{\bf s}\right)\sigma_1\\
			& = {\gamma_{\bf s}\over1+\gamma_{\bf s}}\left( {1\over 4\gamma_{\bf s}^2}\sigma_0+ {1\over 2\gamma_{\bf s}}\sigma_1\rho_{\bf s}\sigma_1+  {1\over 2\gamma_{\bf s}}\rho_{\bf s}+\rho_{\bf s}\sigma_1\rho_{\bf s}\sigma_1\right).
		\end{split}
	\end{equation}
	We have $\Tr(\sigma_0)=2$, $\Tr(\sigma_1\rho_ {\bf s} \sigma_1)=\Tr(\rho_s)=1$ and $\Tr(\rho_{\bf s}\sigma_1\rho_{\bf s}\sigma_1)=(1+s_1^2-s_2^2)/2$, but $1/2\gamma_ {\bf s}^2=(1-s_1^2-s_2^2)/2$, so $\Tr(\rho_{\bf s}\sigma_1\rho_{\bf s}\sigma_1)=1/2\gamma_ {\bf s}^2 + s_1^2$ thus, using eq. \eqref{eq:matentries} we have:
	\begin{equation}
		\begin{split}
			\left[\xi(\rho_{\bf s}^{1/2})\right]_{11} 
			&={\gamma_{\bf s}\over2(1+\gamma_{\bf s})}\left({1\over 2\gamma_{\bf s}^2} + {1\over \gamma_{\bf s}}+ {1\over 2\gamma_{\bf s}^2} + s_1^2\right)\\
			&={\gamma_{\bf s}\over2(1+\gamma_{\bf s})}\left( \frac{1+\gamma_ {\bf s}+\gamma_ {\bf s}^2 s_1^2}{\gamma_ {\bf s}^2}\right)\\
			& = {1\over 2\gamma_{\bf s}}\left({1\over 1+\gamma_{\bf s}}  +{\gamma_{\bf s}\over 1+\gamma_{\bf s}}+ {\gamma_{\bf s}^2\over 1+\gamma_{\bf s}}s_1^2\right)={1\over 2\gamma_{\bf s}}\left(1+{\gamma_{\bf s}^2\over 1+\gamma_{\bf s}}s_1^2\right).
		\end{split}
	\end{equation}
	Proceeding in the same way for the remaining coefficients, one arrives to eq. \eqref{eq:xinonpure}. \qed 
\end{enumerate}

We can now state the main result of this section. 

\begin{corollary}
	Let $\bf s$ and ${\bf e}=(e_0,e_1,e_2)$ be a state and an effect of Hering's rebit parameterized by the vectors ${\bf v_s}=(s_1,s_2)$ and ${\bf v_e}=(e_1/e_0,e_2/e_0)$, ${\bf v_s}, {\bf v_e}\in \mathcal D$, respectively. Then, the Lüders operation 
	\beq
	\begin{array}{cccl}
		\chi \circ \psi_{\bf e} \circ \chi^{-1}: & \mathcal S(\spinf) & \longrightarrow &  \tilde{\mathcal S}(\spinf) \\
		&  \frac{1}{2} \begin{pmatrix}
			1 \\ \bf v_s
		\end{pmatrix} & \longmapsto         &  \chi(\psi_{\bf e}({\bf s}))
	\end{array}
	\eeq
	associated to the effect $\bf e$ can be represented as follows.
	\begin{enumerate}
		\item If $\bf s$ is a pure state, then:
		\beq\label{eq:chipsipure}
		\chi(\psi_{\bf e}({\bf s})) = e_0\begin{pmatrix} 1 & {\bf v}_{\bf e}^t \\
			{\bf v}_{\bf e} & {\bf v}_{\bf e}{\bf v}_{\bf e}^t
		\end{pmatrix} \frac{1}{2} \begin{pmatrix}
			1 \\ \bf v_s
		\end{pmatrix}
		\eeq 
		and
		\beq\label{eq:chiphipure}
		\chi(\varphi_{\bf e}({\bf s})) = \frac{1}{1+{\bf v_e}\cdot {\bf v_s}}\begin{pmatrix} 1 & {\bf v}_{\bf e}^t \\
			{\bf v}_{\bf e} & {\bf v}_{\bf e}{\bf v}_{\bf e}^t
		\end{pmatrix} \frac{1}{2} \begin{pmatrix}
			1 \\ \bf v_s
		\end{pmatrix}.
		\eeq 
		\item If $\bf s$ is a mixed state, then:
		\begin{equation}\label{eq:chipsimixed}
			\chi(\psi_{\bf e}({\bf s}))={e_0\over \gamma_ {\bf v_e}}B({\bf v}_{\bf e}) \, \frac{1}{2} \begin{pmatrix}
				1 \\ {\bf v_s}
			\end{pmatrix} = \|{\bf e}\|_{\mathcal M}B({\bf v}_{\bf e}) \, \frac{1}{2} \begin{pmatrix}
				1 \\ {\bf v_s}
			\end{pmatrix} ,
		\end{equation}
		and 
		\begin{equation}\label{eq:chiphimixed}
			\chi(\varphi_{\bf e}({\bf s}))=
			{1\over \gamma_ {\bf v_e}(1+{\bf v}_{\bf e}\cdot{\bf v}_{\bf s})}B({\bf v}_{\bf e}) \, \frac{1}{2} \begin{pmatrix}
				1 \\ {\bf v_s}
			\end{pmatrix},
		\end{equation}
		where
		\beq
		\gamma_{\bf v_e}=\frac{1}{\sqrt{1-\|{\bf v}_ {\bf e}\|^2}} = \frac{e_0}{\sqrt{e_0^2-e_1^2-e_2^2}} = \frac{e_0}{\|{\bf e}\|_{\mathcal M}},
		\eeq 
		and
		\beq
		[B({\bf v_e})]=\begin{pmatrix}
			\gamma_{\bf v_e} & \gamma_{\bf v_e}{\bf v}_{\bf e}^t \\\gamma_{\bf v_e}{\bf v}_{\bf e} & \sigma_0 +{\gamma_{\bf v_e}^2\over 1+\gamma_{\bf v_e}}{\bf v}_{\bf e}{\bf v}_{\bf e}^t
		\end{pmatrix}.
		\eeq 
	\end{enumerate}
\end{corollary}

\proof Equations \eqref{eq:chipsipure} and \eqref{eq:chipsimixed} can be obtained by considering eq. \eqref{eq:chipsi} and replacing $\chi(\rho_{\bf s})$ by the vector in eq. \eqref{eq:chirhos} and then by replacing $\xi(\rho_{\bf s}^{1/2})$ with the expressions obtained in eqs. \eqref{eq:xipure}, \eqref{eq:xinonpure} for the case of pure and mixed states, respectively. 

Equations \eqref{eq:chiphipure}, \eqref{eq:chiphimixed} follow directly from eqs. \eqref{eq:chipsipure} and \eqref{eq:chipsimixed} by using eqs. \eqref{eq:expes}, \eqref{eq:phies}.
\qed

\subsection{Quantum measurements and the Einstein-Poincaré relativistic addition law}\label{subsec:EP}

Given two arbitrary vectors ${\bf u}$ and ${\bf v}$ of $\mathbb R^2$, with $\Vert {\bf u}\Vert <1$ and $\Vert {\bf w}\Vert \leq 1$, their Einstein-Poincaré relativistic sum is defined by:
\begin{equation}\label{eq:reladd}
	{\bf u}\oplus{\bf w}={1\over 1+{\bf u}\cdot{\bf w}}\left\{{\bf u}+{1\over \gamma_{\bf u}}{\bf w}+{\gamma_{\bf u}\over 1+\gamma_{\bf u}}({\bf u}\cdot{\bf w}){\bf u}\right\},
\end{equation}
with
\begin{equation}
	\gamma_{\bf u}={1\over \sqrt{1-\Vert {\bf u}\Vert^2}},
\end{equation}
moreover, when $\Vert {\bf u}\Vert \to 1$, $\gamma_{\bf u}\to +\infty$ and the Einstein-Poincaré relativistic sum tends to $\bf u$, so the definition above can be extended also to the case $\|{\bf u}\|=1$ by setting:  ${\bf u}\oplus{\bf w}={\bf u}$.

The following proposition shows that when a visual scene in a state ${\bf s}$ is subjected to the measurement of a perceived color identified with the effect ${\bf e}$, the post-measurement state $\varphi_{\bf e}(\bf s)$ is parameterized by a very  special vector: the relativistic sum between $\textbf{v}_ {\bf e}$ and $\textbf{v}_ {\bf s}$.

\begin{proposition}\label{prop:reladd}
	Let ${\bf e}$ and ${\bf s}$ be an effect and a state of Hering's rebit with associated vectors ${\bf v}_{\bf e}$ and ${\bf v}_{\bf s}$, respectively, i.e.
	\beq 
	\chi(\eta_{\bf{e}})={e_0}\begin{pmatrix}
		1 \\ {\bf v_e}
	\end{pmatrix} 
	\quad \text{and} \quad 
	\chi(\rho_{\bf{s}})=\frac{1}{2}\begin{pmatrix}
		1 \\ {\bf v_s}
	\end{pmatrix}
	\eeq 
	and let $\varphi_{\bf e}({\bf s})$ and  $\psi_{\bf e}({\bf s})$ the post-measurement state and generalized density matrix, respectively. Then:
	\begin{equation}\label{eq:reladd1}
		\chi(\varphi_{\bf e}({\bf s}))=\frac{1}{2}\begin{pmatrix}
			1 \\ {\bf v_e} \oplus {\bf v_s}
		\end{pmatrix}
	\end{equation}
	and:
	\begin{equation}\label{eq:reladd2}
		\chi(\psi_{\bf e}({\bf s}))=e_0(1+{\bf v}_{\bf e}\cdot {\bf v}_{\bf s}) \, \frac{1}{2}\begin{pmatrix}1 \\{\bf v}_{\bf e}\oplus{\bf v}_{\bf s}\end{pmatrix},
	\end{equation}
	where ${\bf v}_{\bf e}\oplus {\bf v}_{\bf s}$ denotes the Einstein-Poincaré sum of ${\bf v}_{\bf e}$ and ${\bf v}_{\bf s}$, which is equal to $\bf v_e$ when $\|{\bf v_e}\|=1$.

\end{proposition}

\proof By a direct computation, it can be checked that, for all ${\bf u}$ such that $\|{\bf u}\|<1$ and ${\bf w}\in \mathcal D$, the following equality holds: 
\begin{equation}
	B({\bf u})\begin{pmatrix}1 \\{\bf w}\end{pmatrix}=\gamma_{\bf u}(1+{\bf u}\cdot{\bf w})\begin{pmatrix}1 \\{\bf u}\oplus{\bf w}\end{pmatrix},
\end{equation}
where
\begin{equation}
	\left[B({\bf u})\right]=\begin{pmatrix}\gamma_{\bf u} & \gamma_{\bf u}{\bf u}^t \\\gamma_{\bf u}{\bf u} & \sigma_0 +{\gamma_{\bf u}^2\over 1+\gamma_{\bf u}}{\bf u}{\bf u}^t\end{pmatrix}.
\end{equation}
By eq. (\ref{eq:chiphimixed}), we have:
\begin{equation}
	\chi(\varphi_{\bf e}({\bf s}))={1\over \gamma_{\bf v_e}(1+{\bf v_e}\cdot{\bf v_s})}B({\bf v}_{\bf e}) \, \frac{1}{2} \begin{pmatrix}1 \\{\bf v}_{\bf s}\end{pmatrix}= \frac{1}{2}\begin{pmatrix}1 \\{\bf v}_{\bf e}\oplus{\bf v}_{\bf s}\end{pmatrix},
\end{equation}
and so, by using eqs. \eqref{eq:expes}, \eqref{eq:phies} and the linearity of $\chi$ we obtain: 
\begin{equation}
	\chi(\psi_{\bf e}({\bf s}))=e_0(1+{\bf v}_{\bf e}\cdot {\bf v}_{\bf s}) \, \frac{1}{2} \begin{pmatrix}1 \\{\bf v}_{\bf e}\oplus{\bf v}_{\bf s}\end{pmatrix}.
\end{equation}
Let us now treat the case $\|{\bf v_e}\|=1$. For all ${\bf v_s}\in \mathcal D$, the following equality can be checked directly:
\beq
\begin{pmatrix}
	1 & {\bf v_e}^t\\ {\bf v_e} & {\bf v_e} {\bf v_e}^t 
\end{pmatrix} \begin{pmatrix}
	1 \\ {\bf v_s}
\end{pmatrix} = (1+{\bf v_e}\cdot {\bf v_s}) \begin{pmatrix}
	1 \\ {\bf v_e}
\end{pmatrix},
\eeq 
hence, by \eqref{eq:chiphipure}, we obtain:
\beq
\chi(\varphi_{\bf e}({\bf s}))=\frac{1}{2} \begin{pmatrix}
	1 \\ {\bf v_e}
\end{pmatrix}= \frac{1}{2} \begin{pmatrix}
	1 \\ {\bf v_e} \oplus {\bf v_s}
\end{pmatrix}
\eeq 
which leads to:
\begin{equation}
	\chi(\psi_{\bf e}({\bf s}))=e_0(1+{\bf v}_{\bf e}\cdot {\bf v}_{\bf s}) \, \frac{1}{2} \begin{pmatrix}1 \\{\bf v}_{\bf e}\end{pmatrix} = e_0(1+{\bf v}_{\bf e}\cdot {\bf v}_{\bf s}) \, \frac{1}{2}\begin{pmatrix}1 \\{\bf v}_{\bf e}\oplus{\bf v}_{\bf s}\end{pmatrix}.
\end{equation}
\qed 

\textit{Remark}: the proposition just proven provides an alternative argument, with respect to \cite{Kim:11}, of the fact that the Einstein-Poincaré gyrogroup of the unit disk is isomorphic, as a gyrogroup, to the gyrogroup of the space of density matrices of the rebit, since the composition law on the latter space can be represented by the post-measurement state $\varphi_{\bf e}({\bf s}).$
\section{Applications: a chromatic  matching equation and the quantitative description of Hunt's effect}\label{sec:chromaticmatch}

In this section we will show relevant applications of the results obtained in the previous section. Specifically, we will provide an equation to  characterize chromatic matching and we will  describe quantitatively the well-known Hunt's effect.

In order to do that in a meaningful way, we first need to adapt to the notations of the present paper the mathematical vocabulary introduced in  \cite{Berthier:21JMP} to formalize Yilmaz's heuristic analogy between color perception and the special theory of relativity, see  \cite{Yilmaz:62} and \cite{Prencipe:20}.

Let us consider an effect ${\bf e}=(e_0,e_1,e_2)$ and compute its expectation value on the achromatic state ${\bf s_a}$ described by the density matrix $\rho_{\bf 0}$ that maximizes the von Neumann entropy. Since $\bf v_{s_a}=0$, thanks to eq. \eqref{eq:expes} we find:
\beq
\langle {\bf e} \rangle_{\bf s_a} = e_0.
\eeq 
Since in the achromatic state no chromatic information is available, it is natural to associate $e_0$, the first component of an effect,  to the achromatic information of the perceived color corresponding to the effect $\bf e$. This quantity in literature is referred to as luminance, brightness, intensity and so on. To avoid a misleading use of one of these terms, in \cite{Berthier:21JMP} the word `magnitude' has been preferred and this is also the term that will be employed  here.  

The chromatic information of the perceived color described by $\bf e$ is contained in the remaining components, i.e. in the effect vector ${\bf v_e}=(e_1/e_0,e_2/e_0)\in \mathcal D$.

Empirical evidence shows us that, when a human being observes for a sufficient amount of time a color stimulus represented by a broadband light, or the surface of a material with non-selective reflective properties illuminated by it, then the color sensation experienced is reported to be achromatic, i.e. with zero saturation. In this situation, the observer, denoted with $o$, is said to be \textit{adapted} to the state $\bf s$ in which the visual scene has been prepared and the effect $\bf e$ identified by the perceived color registered has ${\bf v_ e=\bf 0}$, i.e. $e_1=e_2=0$.

\subsection{The chromatic matching equation}\label{subsec:chromaticmatchingeq}
Let us consider a visual scene prepared in the achromatic state $\bf s=s_a$, so that $\bf v_{s_a}=0$, $\rho_{\bf s_a}=\rho_{\bf 0}$ and:
\beq
\chi(\rho_{\bf 0})=\frac{1}{2} \begin{pmatrix}
	1\\ {\bf 0}
\end{pmatrix},
\eeq  
then, by the definition given above, the only perceived color that an adapted observer can  register in this situation is an effect $\bf e_a$ such that $\bf v_{e_a}=0$. From eq. \eqref{eq:reladd1} it follows that:
\begin{equation}
	\chi(\varphi_{{\bf e_a}}({\bf s}))=\frac{1}{2}\begin{pmatrix}1 \\{\bf v}_{{\bf s_a}}\end{pmatrix} = \frac{1}{2} \begin{pmatrix}1 \\{\bf 0}\end{pmatrix},
\end{equation}
so, by applying $\chi^{-1}$ to both sides of the previous equation, we get $\varphi_{{\bf e_a}}({\bf s_a})=\rho_{\bf 0}$, which can be written in an even more evocative way by identifying $\rho_{\bf 0}$ with ${\bf s_a}$ itself:
\beq\label{eq:idLuders}
\varphi_{{\bf e_a}}({\bf s_a})={\bf s_a},
\eeq
i.e. the achromatic state is invariant with respect to measurements of adapted observers.

In spite of its simplicity, this result has an important consequence: it permits to  meaningfully discuss the \textit{color measurement of an adapted observer} in a visual scene prepared in any state $\bf s$. In fact, it is sufficient to integrate the adaptation process in the preparation of the state $\bf s$ via a so-called `sequential operation', see for instance \cite{Busch:97}. 

Firstly, the scene is prepared in the achromatic state $\bf s_a$ and the observer has the time to adapt by observing for a sufficient amount of time the  achromatic stimulus, when adaptation is complete, the visual scene will still be in the state $\bf s_a$. 

Secondly, the scene is prepared in a state $\bf s$ by suddenly modifying  the color stimulus, thus allowing the adapted observer to perform a color measurement that, this time, will induce a non-trivial operation dependent on the effect $\bf e$ that the adapted observer registers as perceived color. 

Since the first state transformation is the identity, we can simply describe mathematically the sequential operation as \textit{the state change from $\bf s$ to $\varphi_{\bf e}(\bf s)$ induced by the measurement of the perceived color $\bf e$ performed by an adapted observer}.

Having defined precisely this important concept and thanks to proposition \ref{prop:reladd}, we can now state rigorously the main result of this section about a fundamental equation of colorimetry.

\begin{proposition}[Chromatic matching equation]\label{prop:chromatch}
	Consider $({\bf e}^1,{\bf s}_1)$ and $({\bf e}^2,{\bf s}_2)$, where ${\bf s}_1$ and ${\bf s}_2$ are states representing the preparations of a visual scene and ${\bf e}^1$ and ${\bf e}^2$ are effects representing the perceived colors identified by two different observers $o_1$ and $o_2$, adapted to the state ${\bf s}_1$ and ${\bf s}_2$, respectively.
	Then, 
	\begin{equation}
		\varphi_{{\bf e}^1}({\bf s}_1)=\varphi_{{\bf e}^2}({\bf s}_2),
	\end{equation}
	or, equivalently\footnote{We have: ${\bf v}_{{\bf e}^1}=(e_1^1/e_0^1,e_2^1/e_0^1)$, ${\bf v}_{{\bf e}^2}=(e_1^2/e_0^2,e_2^2/e_0^2)$, which explains why, in this proposition, effects have upper indices.}, 
	\begin{equation}
		{\bf v}_{{\bf e}^1}\oplus{\bf v}_{{\bf s}_1}={\bf v}_{{\bf e}^2}\oplus{\bf v}_{{\bf s}_2},
	\end{equation}
	is the chromatic matching equation between $({\bf e}^1,{\bf s}_1)$ and $({\bf e}^2,{\bf s}_2)$ that establishes the perception of the same  chromatic information.
\end{proposition}

\subsection{The Hilbert-Klein distance as a chromaticity constancy metric}\label{subsec:HilbertKlein}

As we have seen, chromatic vectors live in the unit disk
\begin{equation}
	\mathcal D=\left\{{\bf v}\in \mathbb R^2,\ \Vert {\bf v}\Vert\leq 1\right\},
\end{equation}
the aim here is to investigate the metric properties of $\mathcal D$ in relation with the Einstein-Poincaré relativistic addition law obeyed by chromatic vector in the parameterization of  post-measurement state changes.

The results that we are going to discuss next constitute an adaptation to our color perception model of analogous results obtained by A.A.  Ungar in the context of special relativity. We refer the reader also to \cite{Fock:59}, page 39, for a proof based on calculus of variations, and to \cite{Berthier:21JMP} for a simple proof in the special case of collinear vectors.

Given a vector ${\bf v}$ in $\mathcal D$, we denote $\ominus {\bf v}$ the Einstein-Poincaré opposite of ${\bf v}$, i.e. the vector that, for all ${\bf u}\in \mathcal D$, satisfies:
\begin{equation}
	{\bf u}\ominus {\bf v}={\bf u}\oplus(-{\bf v}).
\end{equation}
A particularly interesting distance between ${\bf u},{\bf v} \in \mathcal D$ is defined as follows:
\begin{equation}\label{eq:Ungar}
	d({\bf u},{\bf v})=\Vert {\bf u}\ominus{\bf v}\Vert.
\end{equation}
Following Ungar, see \cite{Ungar:08} page 250, the line element
\begin{equation}
	ds^2=\Vert {\bf v}\ominus({\bf v}+d{\bf v})\Vert^2
\end{equation}
coincides with the Riemannian line element
\begin{equation}
	ds^2= {(1-v_2^2)dv_1^2\over (1-\Vert {\bf v}\Vert^2)^2}+{2v_1v_2dv_1dv_2\over (1-\Vert {\bf v}\Vert^2)^2}+{(1-v_1^2)dv_2^2\over (1-\Vert {\bf v}\Vert^2)^2},
\end{equation}
that is with the line element of the  Hilbert-Klein metric $d_{\mathcal K}$ of $\mathcal D$. This implies, see \cite{Fock:59} or \cite{Ungar:08}, that
\begin{equation}
	d_{\mathcal K}({\bf u},{\bf v})=\tanh^{-1}\Vert -{\bf u}\oplus{\bf v}\Vert.
\end{equation}
As a consequence, we have the following proposition.
\begin{proposition}[Chromaticity constancy]
	Given $({\bf e}^1,{\bf s}_1)$ and $({\bf e}^2,{\bf s}_2)$ as in proposition \ref{prop:chromatch}, the chromatic matching equation
	\begin{equation}
		\varphi_{{\bf e}^1}({\bf s}_1)=\varphi_{{\bf e}^2}({\bf s}_2) \iff 
		{\bf v}_{{\bf e}^1}\oplus{\bf v}_{{\bf s}_1}={\bf v}_{{\bf e}^2}\oplus{\bf v}_{{\bf s}_2},
	\end{equation}
	implies 
	\begin{equation}\label{eq:minus}
		{\bf v}_{{\bf s}_1}\ominus{\bf v}_{{\bf s}_2}=-{\bf v}_{{\bf e}^1}\oplus{\bf v}_{{\bf e}^2},
	\end{equation}
	and 
	\begin{equation}\label{eq:chromconst}
		d_{\mathcal K}({\bf v}_{{\bf s}_1},{\bf v}_{{\bf s}_2})=d_{\mathcal K}({\bf v}_{{\bf e}^1},{\bf v}_{{\bf e}^2}).
	\end{equation}
\end{proposition}
This last equation shows that {\em the Hilbert-Klein metric on $\mathcal D$ is a chromaticity constancy metric}, in the following sense: a chromatic matching implies that the Hilbert-Klein distance between the state chromaticity vectors ${\bf v}_{{\bf s}_1}$ and ${\bf v}_{{\bf s}_2}$ is the same as the one between the effect chromaticity vectors ${\bf v}_{{\bf e}^1}$ and ${\bf v}_{{\bf e}^2}$. The fact that (\ref{eq:minus}) implies (\ref{eq:chromconst}) is not straightforward since it requires some properties of so-called gyrostructures that can be consulted in \cite{Ungar:05}, pages 194-195.

\textit{Remark}. The Hilbert-Klein metric on the disk $\mathcal D$ is a hyperbolic metric whose curvature is $-1$. Contrary to the Poincaré metric, which is also a hyperbolic metric with curvature equal to $-1$, the geodesics of the Hilbert-Klein model are straight segments, and more precisely the chords of $\mathcal D$. We refer the reader to \cite{Berthier:21JMP} for a concise state of the art on the use of hyperbolic metrics in the context of color perception.

\subsection{Hunt's effect}\label{subsec:Hunt}

In the last application that we discuss, we explain now how to recast Hunt's perceptual effect in our mathematical framework. 
Hunt's effect was first explicitly discussed in the 1950 paper \cite{Hunt:50} and it is usually popularized as follows: {\em `The color appearances of objects change significantly when the overall luminance level changes. The Hunt effect can be summarized by the statement that colorfulness of a given stimulus increases with luminance level'}, \cite{Fairchild:13}.

The first problem that we have to face if we want to formalize this sentence is to find a suitable representation of the concept of luminance and colorfulness. The first has been already treated: we have seen that the first coordinate of an effect, its magnitude, can be considered as natural candidate for the luminance. The second is trickier and to analyze it we quote the generally accepted informal definition of colorfulness, see e.g. \cite{Fairchild:13}: `\textit{colorfulness is the attribute of a visual perception according to which the perceived color of an area appears to be more or less chromatic}'. 

In our model, what distinguishes between a `more or less chromatic' color in a given state $\bf s$ is the norm of the corresponding chromatic vector $\bf v_s$: if $\|{\bf v_s}\|=1$, then the color corresponds to a pure hue, and so its colorfulness is maximal, on the contrary the color is achromatic if $\|{\bf v_s}\|=0$. Hence, it seems natural to assume that colorfulness is a monotonically increasing function of $\|{\bf v_s}\|$. As it will be clear in the following, this is the only information that we need in order to prove that our model predicts the Hunt effect, even if the precise analytical identification of the colorfulness function remains a very interesting open problem.

Following this assumption, the formalization of Hunt's effect within our mathematical framework can be represented by a logical implication, namely: given two couples effect-state $({\bf e}^1,{\bf s}_1)$ and $({\bf e}^2,{\bf s}_2)$ as in proposition \ref{prop:chromatch} satisfying, moreover, the condition $0< e^1_0<e^2_0\le 1$, then:
\beq\label{eq:mathHunt}
\varphi_{{\bf e}^1}({\bf s}_1)=\varphi_{{\bf e}^2}({\bf s}_2) \implies \|{\bf v_{s_2}}\|\ge \|{\bf v_{s_1}}\|,
\eeq 
i.e., if the chromatic information registered by $({\bf e}^1,{\bf s}_1)$ matches that of $({\bf e}^2,{\bf s}_2)$, then the colorfulness perceived by the observer adapted to $e_0^2$ on the state ${\bf s}_2$ is larger than that perceived by the observer adapted to $e_0^1$ on the state ${\bf s}_1$.

Let us now show how we can arrive to the implication \eqref{eq:mathHunt}: let us start with two effects ${\bf e}^1=(e^1_0,0,0)$ and ${\bf e}^2=(e^2_0,0,0)$ issued by two adaptations to the achromatic state, with magnitude levels satisfying $0< e^1_0<e^2_0\le 1$, and suppose that a non-achromatic color stimulus is introduced in the visual scene, then the state parameterization passes from $\bf 0$ to a  chromatic vector $\bf v_s\neq 0$. Since ${\bf v}_{{\bf e}^1}={\bf v}_{{\bf e}^2}=0$, eq. \eqref{eq:reladd2} gives:
\begin{equation}\label{eq:twopsi}
	\chi(\psi_{{\bf e}^1}({\bf s}))=e^1_0\, \frac{1}{2}\begin{pmatrix}1 \\{\bf v}_{{\bf s}}\end{pmatrix}
	\quad \text{ and } \quad 
	\chi(\psi_{{\bf e}^2}({\bf s}))=e^2_0\, \frac{1}{2}\begin{pmatrix}1 \\{\bf v}_{{\bf s}}\end{pmatrix}.
\end{equation}
From the condition $0< e^1_0<e^2_0\le 1$ we get $1<e_0^2/e_0^1\le 1/e_0^1$, hence there exists $A>0$ such that $e_0^2=e_0^1(1+A)$, with the constraint
\beq\label{eq:constraintA}
e_0^1\le \frac{1}{1+A},
\eeq 
to guarantee that $e_0^2\le 1$. A representation of the term $A$ that will turn out to be particularly convenient for our analysis can be built thanks to the continuous path of non-null chromatic vectors collinear with $\bf v_s$ defined as follows:
\beq
\begin{array}{cccl}
	{\bf u} : & \left(0,\frac{1}{\|{\bf v}_{\bf s}\|}\right] &  \longrightarrow & \mathcal D \\
	&  \lambda  & \longmapsto         &  {\bf u}(\lambda):=\lambda {\bf v_s},
\end{array}
\eeq
where $\|{\bf u}\|:(0,1/\|{\bf v}_{\bf s}\|] \to (0,1]$, $\|{\bf u}\|(\lambda):=\|{\bf u}(\lambda)\|$ is a strictly increasing function of $\lambda$.

If we set $A\equiv {\bf u}(\lambda) \cdot {\bf v_s} = \lambda \Vert {\bf v}_{\bf s}\Vert^2 >0$, then we can write 
\begin{equation}\label{eq:ezerotwone}
	e^2_0=e^1_0(1+{\bf u}(\lambda)\cdot{\bf v_s})=e^1_0(1+\lambda\Vert {\bf v}_{\bf s}\Vert^2),
\end{equation}
the maximal value taken by $e_0^2$ when written as above can be obtained by introducing $\lambda = 1/\Vert {\bf v}_{\bf s}\Vert$ in eq. \eqref{eq:ezerotwone}, which implies:
\beq 
e^2_0\le e^1_0(1+\lambda\Vert {\bf v}_{\bf s}\Vert).
\eeq 
This makes sense as long as the constraint in eq. \eqref{eq:constraintA} is satisfied, i.e.
\begin{equation}\label{eq:constraintezerone}
	e^1_0\leq {1\over 1+\lambda\Vert {\bf v}_{\bf s}\Vert^2} \qquad \forall \lambda \in (0,1/\Vert {\bf v}_{\bf s}\Vert],
\end{equation}
the right-hand side takes its minimal value when $\lambda$ takes its maximal value, i.e. $\lambda = 1/\Vert {\bf v}_{\bf s}\Vert$, which implies
\beq
e_0^1(1+\Vert {\bf v}_{\bf s}\Vert) \le 1,
\eeq
hence, our reasoning is valid whenever $0< e^1_0<e^2_0\le e_0^1(1+\Vert {\bf v}_{\bf s}\Vert)\le 1$. 

From path of chromatic vectors $\bf u$, we can build the path of iso-magnitude effects with collinear chromatic vectors defined as follows:
\beq
\begin{array}{cccl}
	{\bf e}^{\bf u} : & \left(0,\frac{1}{\|{\bf v}_{\bf s}\|}\right] &  \longrightarrow & \mathcal E^*(\mathcal S(\mathcal A)) \\
	&  \lambda  & \longmapsto         &  {\bf e}^{\bf u}(\lambda):=(e_0^1,{\bf u}(\lambda))=(e_0^1,\lambda {\bf v_s}),
\end{array}
\eeq
Eqs. \eqref{eq:reladd2} and \eqref{eq:ezerotwone} imply
\begin{equation}\label{eq:psitilde1}
	\chi(\psi_{{\bf e}^{{\bf u}}(\lambda)}({\bf s}))=e^1_0(1+{\bf u}(\lambda)\cdot {\bf v_s}) \, \frac{1}{2} \begin{pmatrix}1 \\{\bf u}(\lambda)\oplus{\bf v}_{{\bf s}}\end{pmatrix}=e^2_0 \, \frac{1}{2} \begin{pmatrix}1 \\{\bf u}(\lambda)\oplus{\bf v}_{{\bf s}}\end{pmatrix},
\end{equation}
thus showing the usefulness of writing the magnitude $e_0^2$ as in eq. \eqref{eq:ezerotwone}.  

Now, if we define $\widetilde{\bf s}(\lambda)$ to be the state with chromatic vector ${\bf v}_{\tilde{\bf s}(\lambda)}:={\bf u}(\lambda)\oplus{\bf v}_{{\bf s}}$, then, from the second formula in eq. \eqref{eq:twopsi} we get:
\beq\label{eq:psitilde2}
\chi(\psi_{{\bf e}^2}(\tilde{\bf s}(\lambda)))=e^2_0 \, \frac{1}{2} \begin{pmatrix}1 \\{\bf u}(\lambda)\oplus {\bf v}_{{\bf s}}\end{pmatrix},
\eeq 
by comparing eqs. \eqref{eq:psitilde1} and \eqref{eq:psitilde2} we obtain the following chromatic matching:
\begin{equation}
	\varphi_{{\bf e}^{{\bf u}}(\lambda)}({\bf s})=\varphi_{{\bf e}^2}(\widetilde{\bf s}(\lambda)).
\end{equation}
Consequently, the chromatic sensation perceived with magnitude $e^1_0$ corresponding to the effect ${\bf e}^{{\bf u}}(\lambda)$ as the result of a measurement performed on the state ${\bf s}$ is the same as the chromatic sensation perceived with magnitude $e^2_0$ corresponding to the effect ${\bf e}^2$ as the result of a measurement performed on the state $\widetilde {\bf s}(\lambda)$.

Another geometrical way to interpret our construction consist in  considering the continuous path
\begin{equation}
	\begin{array}{cccl}
		\chi_{\bf e}^{\bf u} : & \left[0,1\right] &  \longrightarrow & \tilde {\mathcal{S}}(\R\oplus \R^2) \\
		&  t  & \longmapsto         &  \chi_{\bf e}^{\bf u} (t):=\chi(\psi_{{\bf e}^{t{\bf u}}(\lambda)}({\bf s}))
	\end{array}
\end{equation}
that links the vector identified by $t=0$, i.e.
\begin{equation}
	\chi(\psi_{{\bf e}^1}({\bf s}))={e^1_0\over 2} \, \frac{1}{2} \begin{pmatrix}1 \\{\bf v}_{{\bf s}}\end{pmatrix},
\end{equation}
with the vector identified by $t=1$, i.e.
\begin{equation}
	\chi(\psi_{{\bf e}^{{\bf u}}(\lambda)}({\bf s}))={e^2_0\over 2} \, \frac{1}{2} \begin{pmatrix}1 \\{\bf v}_{\tilde{\bf s}(\lambda)}\end{pmatrix}=\chi(\psi_{{\bf e}^2}(\tilde{\bf s}(\lambda))),
\end{equation}
and such that:
\begin{equation}
	\varphi_{{\bf e}^{{\bf u}}(\lambda)}({\bf s})=\varphi_{{\bf e}^2}({\widetilde {\bf s}(\lambda)})\neq \varphi_{{\bf e}^2}({\bf s}).
\end{equation}

The states $\bf s$ and $\widetilde {\bf s}(\lambda)$ are the equivalent of ${\bf s}_1$ and ${\bf s}_2$, respectively, in the discussion at the beginning of this subsection. Therefore, to prove that our model predicts the Hunt's effect we must just show that $\|{\bf v}_{\widetilde {\bf s}(\lambda)}\|\ge  \|{\bf v}_{\bf s}\|$, which is actually fairly easy. In fact, 
\begin{equation}\label{eq:slambda}
	{\bf v}_{\tilde{\bf s}(\lambda)}={\bf u}(\lambda)\oplus{\bf v}_{\bf s}=\lambda {\bf v_s} \oplus {\bf v_s}={(1+\lambda)\over 1+\lambda\Vert {\bf v_s}\Vert^2}{\bf v}_{\bf s},
\end{equation}
so, if $\Vert{\bf v}_{\bf s}\Vert=1$ we have ${\bf v}_{\tilde{\bf s}(\lambda)}={\bf v_s}$, but if $\Vert{\bf v}_{\bf s}\Vert<1$ then:
\begin{equation}\label{eq:dilationcoeff}
	{1+\lambda\over1+\lambda\Vert{\bf v}_{\bf s}\Vert^2}>1,
\end{equation}
hence, from eq. \eqref{eq:slambda},
\begin{equation}\label{eq:increasedcolfness}
	\Vert {\bf v}_{\tilde{\bf s}(\lambda)} \Vert={(1+\lambda)\over 1+\lambda\Vert {\bf v_s}\Vert^2} \Vert{\bf v}_{\bf s}\Vert >\Vert{\bf v}_{\bf s}\Vert,
\end{equation}
i.e. the colorfulness of the chromatic state $\tilde{\bf s}(\lambda)$ is larger than that of $\bf s$ whenever $\bf s$ is not a pure state, otherwise both $\tilde{\bf s}(\lambda)$ and $\bf s$ are pure.

The colorfulness dilatation coefficient appearing on the left-hand side of inequality \eqref{eq:dilationcoeff} can be more meaningfully  re-expressed in terms of the initial colorfulness $\|{\bf v_s}\|$ and the two magnitude values $e_0^1$ and $e_0^2$. In fact, from eq. \eqref{eq:ezerotwone} we have that $\lambda=(e_0^2-e_0^1)/e_0^1\|{\bf v_s}\|$, hence:
\beq 
\beta(\|{\bf v}_{\bf s}\|,e_0^1,e_0^2)=\frac{1+\lambda}{1+\lambda \|{\bf v_s}\|^2} = \frac{e_0^1}{e_0^2} +  
{e_0^2-e_0^1 \over e_0^2\Vert{\bf v}_{\bf s}\Vert^2}.\eeq 
$\beta$ is a non linear function of its arguments, in particular, regarding its behavior with respect to $\|{\bf v_s}\|$, the second term implies that its dilatation power is much stronger when the colorfulness of $\bf s$ is small, it decreases as the colorfulness of $\bf s$ increases until becoming null when $\bf s$ is pure, which means that  the colorfulness perceived in pure states is invariant with respect to magnitude changes.

We resume and formalize what we have proven in the following proposition, which shows that it is possible to quantify the variation of colorfulness with respect to the variation of magnitude.
\begin{proposition}[Hunt's effect]
	Let ${\bf e}^1=(e_0^1,0,0)$, $e_0^1>0$, be an adapted observer and let $\bf v_s\neq 0$ be a chromatic vector. Let also $e_0^2$  be a second magnitude level satisfying $0< e^1_0<e^2_0\le e_0^1(1+\Vert {\bf v}_{\bf s}\Vert)\le 1$. Then, increasing the magnitude from $e_0^1$ to $e_0^2$ increases the  colorfulness of the colors perceived in the state ${\bf s}$ by the following factor:
	\begin{equation}
		\beta(\|{\bf v}_{\bf s}\|,e_0^1,e_0^2) = \frac{e_0^1}{e_0^2} +  
		{e_0^2-e_0^1 \over e_0^2\Vert{\bf v}_{\bf s}\Vert^2}.
	\end{equation}
\end{proposition}

The case examined so far deals with collinear chromatic vectors. Let us now illustrate with an example what happens if ${\bf u}$ is not collinear with ${\bf v}_{\bf s}$. To make computations as simple as possible, we consider ${\bf u}=(\sqrt3/(2\sqrt2),\sqrt3/(2\sqrt2))$ and ${\bf v}_{\bf s}=(1/\sqrt3,0)$. We  have 
\begin{equation}
	e^2_0=e^1_0(1+{\bf u}\cdot{\bf v}_{\bf s})=e^1_0(1+1/(2\sqrt2)).
\end{equation}
In this case $\gamma_{\bf u}={1\over \sqrt{1-\Vert {\bf u}\Vert^2}}=2$, so, by eq. \eqref{eq:reladd}, 
\begin{equation}
	{\bf u}\oplus{\bf v}_{\bf s}={2\sqrt2\over 2\sqrt2+1}\left\{\begin{pmatrix} \sqrt3/(2\sqrt2)\\\sqrt3/(2\sqrt2)\end{pmatrix}+ {1\over 2} \begin{pmatrix} 1/\sqrt3\\0\end{pmatrix}  +{1\over 3\sqrt2}   \begin{pmatrix} \sqrt3/(2\sqrt2)\\\sqrt3/(2\sqrt2)\end{pmatrix}    \right\},
\end{equation}
i.e.
\begin{equation}
	{\bf u}\oplus{\bf v}_{\bf s}={1\over \sqrt6(1+2\sqrt2)}\begin{pmatrix} 3(1+\sqrt2)\\3\sqrt2+1\end{pmatrix}.
\end{equation}
By direct computation, it can be verified that  (\ref{eq:increasedcolfness}) still holds true, however the vector ${\bf u}\oplus{\bf v}_{\bf s}$ is no longer collinear with ${\bf v}_{\bf s}$, which means that not only colorfulness has increased, but also that the perceived hue has changed.

\section {Discussion and future perspectives}\label{sec:conclusions}

We have introduced some fundamental concepts and tools borrowed from the theory of quantum measurement into the mathematical framework for color perception developed in  \cite{BerthierProvenzi:19,Berthier:2020,Berthier:2021JofImaging,Berthier:21JMIV,Berthier:21JMP}. 

We have justified why we propose to identify the act of observing a color with a quantum unsharp observable and the measurement outcome, i.e. a perceived color, with an effect.

Effects are then shown to be extremely fit to describe color perception, as they permit to naturally restrict the infinite cone of perceptual colors to a finite double convex cone which is in  perfect agreement with the geometrical structure guessed by Ostwald and deValois. 

The analysis of state transformation with the formalism of Lüders transformations and Kraus operators led us to the main theoretical results of our paper: first, it is possible to relate post-measurement transformations with Lorentz boosts and the Einstein-Poincaré relativistic addition of chromatic vectors; secondly, the Hilbert-Klein metric on the unit disk can be used to express chromaticity constancy.

The final part of the paper is devoted to applications: we have derived the fundamental law of chromatic matching and applied it to study the well-known Hunt effect. 

The future perspectives that we envisage are at least three. The first, and probably the easiest one, consists in the use of eq. \eqref{eq:chiphimixed} obtained here to improve the performances of the white balance algorithm based on Lorentz boosts proposed in \cite{Guennec:21}. 

The second consists in trying to find rigorous mathematical analogues of the many colorimetric concepts introduced by the CIE in the context of the framework described in this paper. Concepts as lightness, brightness, colorfulness, chroma, purity, saturation, etc. are defined in a quite elusive way. We have already quoted the standard definition of colorfulness, those of chroma and saturation are no less involved and circular: `\textit{Chroma is the colorfulness of an area judged as a proportion of the brightness of a similarly illuminated area that appears white or highly transmitting. Saturation is the colorfulness of an area judged in proportion to its brightness}', \cite{Fairchild:13}. Speaking of color characteristics in words in not an easy task, as incisively pointed out by Wittgenstein in \cite{Wittgenstein:77}: {\em `when we're asked `What do the words red, blue, black, white mean?' we can, of course, immediately point to things which have these colours, but our ability to explain the meanings of these words goes no further! For the rest, we have either no idea at all of their use, or a very rough and to some extent false one. The word whose meaning is not clear is pure or saturated. How do we learn its meaning? How can we tell if people mean the same thing by it?'}

We believe that the use of the mathematical concepts and results defined in this paper can lead towards the formalization of colorimetric attributes. A strong indication that this strategy may work is given by the fact that we have already provided a formalization of the concepts of hue, saturation and magnitude and, remarkably, that our analysis of the Hunt effect allowed us to quantify by means of precise mathematical equations the way perceived colors evolve with respect to experimental conditions.

The third perspective is the most ambitious one and consists in extending our results, which refer to isolated colors, to composite system describing color in context. The simplest example is given by a central color stimulus embedded in a uniform colored background: it is well-known that variations of the background modify the perception of the central color. It seems reasonable to try to address this problem using Stinespring's theorem applied on a composite system of two rebits \cite{Heinosaari:2011}, with the background playing the role of the ancillary state.



\begin{thebibliography}{100}
	
	\bibitem{Arrighi:04}
P.~ Arrighi and C.~Patricot.
On quantum operations as quantum states.
{\em Annals of Physics}, 311(1):26--52, 2004.

\bibitem{Baez:12}
J.~C. Baez.
Division algebras and quantum theory.
{\em Foundations of Physics}, 42(7):819--855, 2012.

\bibitem{Berthier:2020}
M.~Berthier.
Geometry of color perception. {P}art 2: perceived colors from real
quantum states and {H}ering’s rebit.
{\em The Journal of Mathematical Neuroscience}, 10(1):1--25, 2020.

\bibitem{Berthier:21JMP}
M.~Berthier, V.~Garcin, N.~Prencipe, and E.~Provenzi.
The relativity of color perception.
{\em {J}ournal of {M}athematical {P}sychology}, 103, 102562, 2021.

\bibitem{Berthier:21JMIV}
M.~Berthier and E.~Provenzi.
From {R}iemannian trichromacy to quantum color opponency via
hyperbolicity.
{\em {J}ournal of {M}athematical {I}maging and {V}ision}, 
63(6), 681-688, 2021.

\bibitem{BerthierProvenzi:19}
M.~Berthier and E.~Provenzi.
When geometry meets psycho-physics and quantum mechanics: {M}odern
perspectives on the space of perceived colors.
In {\em GSI 2019}, volume 11712 of {\em Lecture Notes in Computer
	Science}, pages 621--630. Springer Berlin-Heidelberg, 2019.

\bibitem{Berthier:2021JofImaging}
M.~Berthier and E.~Provenzi.
The quantum nature of color perception: Uncertainty relations for
chromatic opposition.
{\em Journal of Imaging}, 7(2):40, 2021.

\bibitem{Buchsbaum:83}
G.~Buchsbaum and A.~Gottschalk.
Trichromacy, opponent colours coding and optimum colour information
transmission in the retina.
{\em Proc. Royal Society of London B}, 220:89--113, 1983.

\bibitem{Busch:97}
P.~Busch, M.~Grabowski, and P.~J Lahti.
{\em Operational quantum physics}, volume~31.
Springer Science \& Business Media, 1997.

\bibitem{Busch:16}
P.~Busch, P.~Lahti, J.-P.~Pellonp{\"a}{\"a}, and K.~Ylinen.
{\em Quantum measurement}, volume~23.
Springer, 2016.

\bibitem{Cohen:85}
J.~B. Cohen and W.~E.~ Kappauf.
Color mixture and fundamental metamers: Theory, algebra, geometry,
application.
{\em The American journal of psychology}, pages 171--259, 1985.

\bibitem{Cie:95}
CIE~Technical Committee et~al.
CIE 117-1995 discomfort glare in interior lighting, 1995.

\bibitem{Davies:70}
E.~B. Davies and J.~T.~ Lewis.
An operational approach to quantum probability.
{\em Communications in Mathematical Physics}, 17(3):239--260, 1970.

\bibitem{deValois:2000}
K.~K.~ de~Valois.
{\em Seeing}.
Academic Press, 2000.

\bibitem{deValois:97}
R.~L.~ de~Valois and K.~K.~de~Valois.
Neural coding of color.
1997.

\bibitem{Dubois:09}
E.~Dubois.
The structure and properties of color spaces and the representation
of color images.
{\em Synthesis Lectures on Image, Video, and Multimedia Processing},
4(1):1--129, 2009.

\bibitem{Emch:2009}
G.G.~Emch
{\em Algebraic methods in statistical mechanics and quantum field theory}.
Courier Corporation, 2009.

\bibitem{Fairchild:13}
M.D.~Fairchild, {\em Color appearance models}. John Wiley \& Sons, 2013.

\bibitem{Faraut:94}
J.~Faraut and A.~Koranyi.
{\em Analysis on Symmetric Cones}.
Clarendon Press, Oxford, 1994.

\bibitem{Fock:59}
V.~Fock.
{\em The theory of space time and gravitation}.
Pergamon Press, 1959.

\bibitem{Granville:94}
W.~C Granville.
The color harmony manual, a color atlas based on the {O}stwald color
system.
{\em Color Research \& Application}, 19(2):77--98, 1994.

\bibitem{Guennec:21}
A.~Guennec, N.~Prencipe and E.~Provenzi. \textit{Color correction with Lorentz boosts}.	  
{\em The 4th International Conference on Image and Graphics Processing}, 162–168, Sanya, China, 2021.

\bibitem{Haag:64}
R.~ Haag and D.~ Kastler.
An algebraic approach to quantum field theory.
{\em Journal of Mathematical Physics}, 5(7):848--861, 1964.

\bibitem{Heinosaari:2011}
T.~Heinosaari and M.~Ziman.
{\em The mathematical language of quantum theory: from uncertainty to
	entanglement}.
Cambridge University Press, 2011.

\bibitem{Hering:1878}
E.~ Hering.
{\em Zur {L}ehre vom {L}ichtsinne: sechs {M}ittheilungen an die
	{K}aiserl. {A}kademie der {W}issenschaften in {W}ien}.
C. Gerold's Sohn, 1878.

\bibitem{Hubel:95}
D.H.~ Hubel.
{\em Eye, Brain, and Vision}.
Scientific American Library, 1995.

\bibitem{Hunt:50}
R. W. G.~Hunt, The effects of daylight and tungsten light-adaptation on color perception, \textit{JOSA}, 40(6): 362--371, 1950.

\bibitem{Jordan:34}
P.~Jordan, J.~von~Neumann, and E.~Wigner.
On an algebraic generalization of the quantum mechanical formalism.
{\em Annals of Math.}, 35:29--64, 1934.

\bibitem{Kim:11}
S.~ Kim.
Distances of qubit density matrices on bloch sphere.
{\em Journal of mathematical physics}, 52(10):102303, 2011.

\bibitem{Koenderink:10}
J.~J,~ Koenderink.
{\em Color for the Sciences}.
The MIT Press, 2010.

\bibitem{Koenderink:03}
J.~J.~ Koenderink and A.~J.~ van Doorn.
{\em Perspectives on colour space}.
Oxford University, 2003.

\bibitem{Krantz:75}
D.~H.~ Krantz.
Color measurement and color theory: I. representation theorem for
grassmann structures.
{\em Journal of Mathematical Psychology}, 12(3):283--303, 1975.

\bibitem{Kraus:83}
K.~Kraus, A.~ B{\"o}hm, J.~D.~ Dollard, and W.H.~Wootters.
States, effects, and operations: fundamental notions of quantum
theory. lectures in mathematical physics at the university of texas at
austin.
{\em Lecture notes in physics}, 190, 1983.

\bibitem{Levinger:80}
B.~W.~Levinger.
The square root of a 2$\times$ 2 matrix.
{\em Mathematics Magazine}, 53(4):222--224, 1980.

\bibitem{Ludwig:64}
G.~Ludwig.
Versuch einer axiomatischen {G}rundlegung der {Q}uantenmechanik und
allgemeinerer physikalischer {T}heorien.
{\em Zeitschrift f{\"u}r Physik}, 181(3):233--260, 1964.

\bibitem{Mccrimmon:1978}
K.~McCrimmon.
Jordan algebras and their applications.
{\em Bulletin of the American Mathematical Society}, 84:612--627,
1978.

\bibitem{Moretti:17}
V.~Moretti.
{\em Spectral theory and quantum mechanics: mathematical foundations
	of quantum theories, symmetries and introduction to the algebraic
	formulation}.
Springer, 2017.

\bibitem{Newton:52}
I.~Newton.
Opticks, or, a treatise of the reflections, refractions, inflections
\& colours of light. {C}ourier {C}orporation.
1952.

\bibitem{Prencipe:20}
N.~Prencipe, V.~Garcin, and E.~ Provenzi.
Origins of hyperbolicity in color perception.
{\em Journal of Imaging}, 6(6), 42, 2020. 

\bibitem{Provenzi:16}
E.~Provenzi, J.~Delon, Y.~Gousseau, and B.~Mazin.
On the second order spatiochromatic structure of natural images.
{\em Vision research}, 120:22--38, 2016.

\bibitem{Provenzi:16modernphysics}
E.~ Provenzi.
A differential geometry model for the perceived colors space.
{\em International Journal of Geometric Methods in Modern Physics},
13(08):1630008, 2016.

\bibitem{Provenzi:20}
E.~ Provenzi.
Geometry of color perception. {P}art 1: {S}tructures and metrics of a
homogeneous color space.
{\em The Journal of Mathematical Neuroscience}, 10(1):1--19, 2020.

\bibitem{Resnikoff:74}
H.L. Resnikoff.
Differential geometry and color perception.
{\em Journal of Mathematical Biology}, 1:97--131, 1974.

\bibitem{Ruderman:98}
D.L. Ruderman, T.W. Cronin, and C.~Chiao.
Statistics of cone responses to natural images: implications for
visual coding.
{\em J. Opt. Soc. Am. A}, 15(8):2036--2045, August 1998.

\bibitem{Russell:16}
S.~J.~ Russell and P.~ Norvig.
{\em Artificial intelligence: a modern approach}.
Malaysia; Pearson Education Limited,, 2016.

\bibitem{Schroedinger:20}
E.~Schr\"odinger.
Grundlinien einer theorie der farbenmetrik im tagessehen ({O}utline
of a theory of colour measurement for daylight vision). {A}vailable in
{E}nglish in {S}ources of {C}olour {S}cience, {E}d. {D}avid {L}. {M}acadam,
{T}he {MIT} {P}ress (1970), 134-82.
{\em Annalen der Physik}, 63(4):397--456; 481--520, 1920.

\bibitem{Strocchi:08}
F.~ Strocchi.
{\em An introduction to the mathematical structure of quantum
	mechanics: a short course for mathematicians}, volume~28.
World Scientific, 2008.

\bibitem{Townsend:85}
P.~K.~Townsend.
The Jordan formulation of quantum mechanics: A review.
Supersymmetry, supergravity, and related topics, 1985.

\bibitem{Ungar:05}
A.~A.~Ungar. Einstein's special relativity: {\em {U}nleashing the power of its hyperbolic geometry}. Computers \& Mathematics with Applications, (2-3), 187-221, 2008.

\bibitem{Ungar:08}
A.~A.~Ungar. Analytic hyperbolic geometry and {A}lbert {E}instein's special theory of relativity. {\em World scientific}, 2008.

\bibitem{vonNeumann:271}
J.~ von~Neumann.
Mathematische {B}egr{\"u}ndung der {Q}uantenmechanik.
{\em Nachrichten von der {G}esellschaft der {W}issenschaften zu
	{G}{\"o}ttingen, {M}athematisch-{P}hysikalische {K}lasse}, 1927:1--57, 1927.

\bibitem{vonNeumann:272}
J.~ von~Neumann.
Wahrscheinlichkeitstheoretischer {A}ufbau der {Q}uantenmechanik.
{\em Nachrichten von der {G}esellschaft der {W}issenschaften zu
	{G}{\"o}ttingen, {M}athematisch-{P}hysikalische {K}lasse}, 1927:245--272,
1927.

\bibitem{Wittgenstein:77}
L.~Wittgenstein, G.~Anscombe, L.~ McAlister and M.~Sch{\"a}ttle. {\em Remarks on colour}, Blackwell Oxford, 1977.

\bibitem{Wootters:2014}
W.~K.~ Wootters.
The rebit three-tangle and its relation to two-qubit entanglement.
{\em Journal of Physics A: Mathematical and Theoretical},
47(42):424037, 2014.


\bibitem{Yilmaz:62}
H.~Yilmaz.
On color perception.
{\em The bulletin of mathematical biophysics}, 24(1), 5--29, 1962.
	
\end{thebibliography}
\end{document}